\documentclass[11pt]{article}
\edef\qedrestoreat{\noexpand\catcode\lq\noexpand\@=\the\catcode\lq\@}
\catcode\lq\@=11

\ifx\protect\undefined\let\protect\relax\fi

\def\qed{\protect\@qed{$\qedsymbol$}}
\def\pushright{\protect\@pushright}

\def\QED{\protect\@qed{{\rm Q.E.D.}}}

\def\QEI{\protect\@qed{{\rm Q.E.I.}}}

\def\Proof{\protect\@Proof}\def\endProof{\protect\@endProof}%

\def\Proofof#1{\protect\@Proofof{#1}}\def\endProofof{\protect\@endProofof}%
\let\proofof\Proofof\let\endproofof\endProofof
\let\proofend\endproofof        

\def\qedsymbol{\raisebox{-.2ex}{$\Box$}}

\def\TheWordProof{\em Proof.}
\def\TheWordProofof#1{\em Proof of #1.}

\def\ProofFont{}

\newif\ifAutoQED\AutoQEDfalse

\newif\ifNumberResults
\ifx\theorem@style\undefined
   \NumberResultstrue
   
\fi

\def\parag@pushright#1{{
    \parfillskip=0pt            
    \widowpenalty=10000         
    \displaywidowpenalty=10000  
    \finalhyphendemerits=0      
    \hbox@pushright             
    #1
    \par}}

\def\hbox@pushright{
    \unskip                     
    \nobreak                    
    \hfil                       
    \penalty50                  
    \hskip.2em                  
    \null                       
    \hfill                      
}%

\newif\if@qed\@qedfalse
\def\save@set@qed{\let\saved@ifqed\if@qed\global\@qedtrue}%
\def\restore@qed{\global\let\if@qed\saved@ifqed}

\def\@Proof{%
   \par\removelastskip\bigskip\penalty100
   \save@set@qed
   \noindent\ProofFont{\TheWordProof\enskip}%
}%
\def\@Proofof#1{%
   \par\removelastskip\bigskip\penalty100
   \save@set@qed
   \noindent\ProofFont{\TheWordProofof{#1}\enskip}%
}%
\def\@endProof{%
   \qed\restore@qed
   \penalty-100 \medskip
}
\def\@endProofof{%
   \qed\restore@qed
   \penalty-100 \medskip
}
\def\@qed#1{%
\if@qed                                 
     \global\@qedfalse
        \ifmmode\ifinner\pushright{#1}
        \else\eqno{\qedsymbol}\fi
        \else\pushright{#1}\fi%
\else\ifhmode\ifinner\else\par\fi\fi
\fi}
\def\@pushright#1{%
  {\ifvmode                             
       \null\hfill{#1}\par              
  \else\ifmmode\maths@pushright{\hbox{#1}}
       \else\ifinner\hbox@pushright{#1}
            \else\parag@pushright{#1}
  \fi  \fi  \fi
}}%
\def\maths@pushright#1{{%
  \ifinner
     \hbox@pushright{#1}%
  \else
     \eqno#1
     \def\]{$$\ignorespaces}
  \fi
}}%
\usepackage{amsmath,amsfonts,amsbsy,amsgen,amscd, amsxtra,amstext,amsopn,amssymb}
\usepackage{latexsym}
\usepackage{bbm}
\usepackage{theorem}
\usepackage{mathrsfs}
\usepackage{epic}
\usepackage{latexsym}
\usepackage{euscript}

\usepackage[all]{xy}

{\theorembodyfont{\slshape}
\newtheorem{theorem}{Theorem}[section]
\newtheorem{proposition}[theorem]{Proposition}
\newtheorem{lemma}[theorem]{Lemma}
\newtheorem{corollary}[theorem]{Corollary}
}
{\theorembodyfont{\rmfamily}
\newtheorem{definition}[theorem]{Definition}

\newtheorem{remark}[theorem]{Remark}
\newtheorem{example}[theorem]{Example}

}

\def\N{\mathbb{N}}
\def\Z{\mathbb{Z}}
\def\R{\mathbb{R}}
\def\Q{\mathbb{Q}}
\def\C{\mathbb{C}}

\def\P{\mathbb{P}}

\def\proj{\mathbb{P}}
\def\Gr{\mathbb{G}}
\def \Ri{\R^\infty}
\def\Ci{{\C^\infty}}

\def \OO{\mathcal{O}}
\def\eZ{{\hat{\Z}}}

\def\mZ{{\mathcal Z}}

\def\Flag{{\mathcal F}}

\def\uF{{\underline{F}}}
\def\bT{{\mathbb{T}}}

\def\Hom{\mathrm{Hom }}

\newcommand{\x}{\times}

\renewcommand{\hat}{\widehat}

\renewcommand{\tilde}{\widetilde}

\def\size{{\rm size}}
\def\codim{{\rm codim}}
\def\Oh{{\cal O}}

\def\BP{{\rm BP}}

\newcommand{\GL}{\mathrm{GL}}

\def\algorithm{\begin{center}
               \begin{minipage}{6in}
               \begin{tabbing}
               \marks}
\def\falgorithm{\end{tabbing}
                \end{minipage}
                \end{center}}
\def\marks{nn\= nn\= nn\= nn\= nn\= nn\= nn\= \kill}

\def\P{{\sf P}}

\def\PSPACE{{\sf PSPACE}}
\def\FPSPACE{{\sf FPSPACE}}

\def\FEXPSPACE{{\sf FEXPSPACE}}

\def\CP{{\#\P}}

\def\GCC{{\sf GCC}}

\def\GapP{{\sf GapP}}

\def\PR{{\rm P}_{\kern-1pt\R}}
\def\PC{{\rm P}_{\kern-1pt\C}}
\def\NPR{{\rm NP}_{\kern-1pt\R}}
\def\NPC{{\rm NP}_{\kern-2pt\C}}
\def\coNPC{{\rm coNP}_{\kern-2pt\C}}
\def\coNPR{{\rm coNP}_{\kern-2pt\R}}
\def\DNPR{{\rm DNP}_{\kern-1pt\R}}
\def\DNPC{{\rm DNP}_{\kern-2pt\C}}
\def\PAR{{\rm PAR}_{\kern-1pt\R}}
\def\PARC{{\rm PAR}_{\kern-1pt\C}}
\def\PHR{{\rm PH}_{\kern-1pt\R}}
\def\PHC{{\rm PH}_{\kern-1pt\C}}
\def\DPHR{{\rm DPH}_{\kern-1pt\R}}
\def\DPHC{{\rm DPH}_{\kern-1pt\C}}

\def\FPR{{\rm FP}_{\kern-1pt\R}}
\def\FPC{{\rm FP}_{\kern-1pt\C}}
\def\FPAR{{\rm FPAR}_{\kern-0.4pt\R}}
\def\FPARC{{\rm FPAR}_{\kern-0.4pt\C}}

\def\CPRi{{\rm \#P}_{\kern-2pt\R}}
\def\CPRd{{\rm D\#P}_{\kern-2pt\R}}
\def\CPCi{{\rm \#P}_{\kern-2pt\C}}
\def\CPCd{{\rm D\#P_{\kern-2pt\C}}}
\def\gCPCi{{\rm \#P}^{\ast}_{\kern-2pt\C}}
\def\gCPRi{{\rm \#P}^{\ast}_{\kern-2pt\R}}
\def\gpr{\preceq_\ast}

\def\FPK{{\rm FP}\kern-1.5pt_K}
\def\FPL{{\rm FP}\kern-1.5pt_L}
\def\CPK{\#{\rm P}\kern-1.5pt_K}
\def\CPL{\#{\rm P}\kern-1.5pt_L}
\def\CHNK{\#\mbox{\sc HN}\kern-1.5pt_K}
\def\CHNL{\#\mbox{\sc HN}\kern-1.5pt_L}
\def\PK{{\rm P}\kern-1.5pt_K}
\def\PL{{\rm P}\kern-1.5pt_L}
\def\NPK{{\rm NP}\kern-1.5pt_K}
\def\NPL{{\rm NP}\kern-1.5pt_L}

\def\FEASR{{\mbox{\sc Feas}_{\kern-0.5pt\R}}}
\def\FEASC{{\mbox{\sc Feas}_{\kern-1pt\C}}}
\def\FEASRbit{{\mbox{\sc Feas}^{\Z}_{\kern-1pt\R}}}
\def\FEASCbit{{\mbox{\sc Feas}^{\Z}_{\kern-1pt\C}}}
\def\HNC{{\mbox{\sc HN}_{\kern-1pt\C}}}
\def\PHNC{{\mbox{\sc PHN}_{\kern-1pt\C}}}
\def\HNCbit{{\mbox{\sc HN}^{\Z}_{\kern-1pt\C}}}
\def\QASC{{\mbox{\sc QAS}_{\kern-1pt\C}}}
\def\QASCbit{{\mbox{\sc QAS}^{\Z}_{\kern-1pt\C}}}
\def\PQAS{{\mbox{\sc ProjQAS}_{\kern-1pt\C}}}
\def\BPQAS{{\mbox{\sc BiProjQAS}_{\kern-1pt\C}}}

\def\DIMR{{\mbox{\sc Dim}_{\kern-0.5pt\R}}}
\def\DIMC{{\mbox{\sc Dim}_{\kern-0.5pt\C}}}
\def\DIMadd{{\mbox{\sc Dim}_{\kern-0.5pt\add}}}
\def\DIMRbit{{\mbox{\sc Dim}^{\Z}_{\kern-0.5pt\R}}}
\def\DIMCbit{{\mbox{\sc Dim}^{\Z}_{\kern-0.5pt\C}}}

\def\REACH{{\mbox{\sc Reach}_{\kern-0.5pt\R}}}
\def\REACHbit{{\mbox{\sc Reach}^{\Z}_{\kern-0.5pt\R}}}
\def\CREACHbit{{\mbox{\sc CReach}^{\Z}_{\kern-0.5pt\R}}}

\def\HILB{{\mbox{\sc Hilbert}}}
\def\HILBbit{{\mbox{\sc Hilbert}}^\Z}
\def\HILBS{{\mbox{\sc Hilbert}_{\rm sm}}}
\def\HILBSbit{{\mbox{\sc Hilbert}^\Z_{\rm sm}}}
\def\Projchar{{\mbox{\sc ProjChar}}}
\def\HIM{{\mbox{\sc HIM}}}
\def\HIMbit{{\mbox{\sc HIM}}^\Z}
\def\GapHNC{{\mbox{\sc $\Delta$HN}_{\kern-1pt\C}}}
\def\GapHNCbit{{\mbox{\sc $\Delta$HN}^{\Z}_{\kern-1pt\C}}}

\def\RankSheaf{{\mbox{\sc RankSheaf}}}
\def\RankSheafbit{{\mbox{\sc RankSheaf}^\Z}}
\def\EulerSheaf{{\mbox{\sc EulerSheaf}}}
\def\EulerSheafbit{{\mbox{\sc EulerSheaf}^\Z}}
\def\coker{{\rm coker}}

\newcommand{\mcal}[1]{\mathcal{#1}}

\newcommand{\cupp}{\smallsmile}
\newcommand{\capp}{\smallfrown}

\def\gGapC{{\mbox{\sc Gap}^\ast_{\kern-2pt\C}}}
\def\gGapR{{\mbox{\sc Gap}^\ast_{\kern-2pt\R}}}
\def\GapC{{\mbox{\sc Gap}_{\kern-1pt\C}}}

\newcommand{\HHNC}{\mbox{\sc HN}_{\kern-1pt\C}}
\newcommand{\PP}{\mathbb{P}}

\newcommand{\trans}{\mathtt{trans}}

\begin{document}
\begin{title}
{\Large {\bf The complexity of computing the Hilbert polynomial of smooth equidimensional complex projective varieties}}
\end{title}
\author{Peter B\"urgisser and Martin Lotz\thanks{Institute of Mathematics,
University of Paderborn, D-33095 Paderborn, Germany.
E-mail: {\tt \{pbuerg,lotzm\}@upb.de}.
Partially supported by DFG grant BU~1371 and Paderborn Institute for Scientific Computation (PaSCo).}
}
\date{\today}
\makeatletter
\maketitle

\begin{abstract}
We continue the study of counting complexity
begun in~\cite{bucu:04, bucl:04a,bucl:04} by proving upper and lower bounds
on the complexity of computing the Hilbert polynomial of a homogeneous ideal.
We show that the problem of computing the Hilbert polynomial of a smooth equidimensional complex
projective variety can be reduced in polynomial time to the problem
of counting the number of complex common zeros of a finite set of multivariate polynomials.
Moreover, we prove that the more general problem of computing the Hilbert polynomial
of a homogeneous ideal is polynomial space hard. This implies polynomial space
lower bounds for both the problems of computing the rank and the Euler characteristic
of cohomology groups of coherent sheaves on projective space, improving the $\CP$-lower bound in Bach~\cite{bach:99}.
\end{abstract}

\section{Introduction}

Despite the impressive progress in the development of algebraic algorithms and computer algebra packages,
the inherent computational complexity of even the most basic problems in algebraic geometry is still
far from being understood.
In \cite{bucu:04} a systematic study of the inherent complexity for computing
algebraic/topological quantities was launched with the goal of characterizing the complexity
of various such problems by completeness results in a suitable hierarchy of complexity classes.
In this article we continue this study by investigating the complexity of computing the Hilbert polynomial
of a complex projective variety $V\subseteq\PP^n$.
This polynomial encodes important information about the variety $V$, like
its dimension, degree and arithmetic genus.

Algorithms for computing Hilbert polynomials were described in ~\cite{momo:83,bicr:91,bast:92}.
Some of these algorithms have been implemented in computer algebra systems and
work quite well in practice. These algorithms are based on the computation of
Gr\"obner bases, which leads to bad upper complexity estimates.
In fact, 
the problem of computing a Gr\"obner basis is exponential space complete \cite{mayr:97}.
Both the cardinality and the maximal degree of a Gr\"obner basis might be doubly exponential in the
number of variables~\cite{mame:82,huyn:86}.
It is generally believed that these bounds are quite pessimistic and that
for problems with ``nice'' geometry, single exponential upper bounds should hold
for Gr\"obner bases. Among the results that are known in this
direction are~\cite{gius:84,difg:91,bamu:93,mayr:97}.
However, currently no upper bound better than exponential space
is known for the computation of the Hilbert function or Hilbert polynomial of a homogeneous ideal.

Based on a lower bound on the homogeneous polynomial ideal membership problem in~\cite{mayr:97}
we are able to show that the problem of computing the Hilbert polynomial is $\FPSPACE$-hard, where
$\FPSPACE$ denotes the complexity class of functions that can be computed in polynomial space
by a Turing machine. As a corollary, we obtain an $\FPSPACE$-lower bound for the problem
of computing the rank of cohomology groups of coherent sheaves
on projective space as well as for the problem of computing the corresponding Euler
characteristic (Corollary~\ref{cor:bettisheaf}), thus improving the $\CP$-lower bound
in Bach~\cite{bach:99}.

The bound on the Castelnuovo-Mumford regularity for the vanishing ideal of
a smooth projective variety in \cite[Thm.~3.12(b)]{bamu:93} suggests that
the computation of the Hilbert polynomial might actually be possible in polynomial
space for {\em smooth} varieties.
The main goal of this article is to prove a stronger result: we show that the problem
$\HILBS$ of computing the Hilbert polynomial of a smooth
equidimensional complex projective variety $V\subseteq\PP^n$ can be reduced in polynomial time to the problem $\#\HNC$
of counting the number of complex common zeros of a finite set of complex multivariate polynomials.
(The input specification for $\HILBS$ involves some subtleties, see \S\ref{se:compl-hilb}.) 
Such reduction can be established in the Turing as well as in the Blum-Shub-Smale model of computation~\cite{blss:89,bcss:95}.
In particular, in the Turing model we obtain an $\FPSPACE$-upper bound for the discrete version of $\HILBS$, 
where the inputs are integer polynomials.

These results are interpreted in the framework of counting complexity.
In \cite{bucu:04} Valiant's counting complexity class $\CP$~\cite{vali:79-2,vali:79-1}
was extended to the framework of
computations over $\C$ in the sense of \cite{bcss:95}. Thus $\CPCi$ is the class of functions from the space
$\Ci$ of finite sequences of complex numbers to $\N\cup\{\infty\}$ which,
roughly speaking, count the number of satisfying witnesses for an input of a problem in $\NPC$.
The problem $\#\HNC$ of counting the number of complex common zeros of a given finite set of
complex polynomials (returning $\infty$ if this number is not finite) turns out to be
complete for the class $\CPCi$.
The main results of \cite{bucu:04,bucl:04} state that both problems to
compute the geometric degree and the topological Euler characteristic
of complex varieties are polynomial time equivalent to $\#\HNC$
(\cite{bucu:04} also contains a corresponding result for the computation of the
Euler characteristic of semialgebraic sets).
Hereby, ``polynomial time equivalent'' is meant in the sense of computations over $\C$.
However, when restricting the inputs to integer coefficient polynomials,
the corresponding discrete problems are also
equivalent in the Turing model of computation (for Turing reductions).
The complexity of these discrete problems is captured by the Boolean part $\GCC$ of $\CPCi$,
which is obtained by restricting the functions in $\CPCi$ to bit strings.
Is is known that $\CP\subseteq\GCC\subseteq\FPSPACE$ \cite{bucu:04}.
One can show along the lines of \cite[\S8.3]{bucu:04} that the problem of computing
the number of connected components of a complex affine algebraic variety
(given by integer coefficients polynomials) is $\FPSPACE$-complete.
This implies that the problem of computing the topological Euler characteristic is strictly
easier than the problem of computing the number of connected components, unless
$\GCC$ collapses with $\FPSPACE$, which we believe to be unlikely.

The class $\CPCi$ captures the complexity of counting the number of solutions to systems of
polynomial equations. It is therefore not surprising that some of the ideas and tools
of intersection theory, enumerative geometry, and Schubert calculus are salient for
our purposes.

Our reduction from $\HILBS$ to $\#\HNC$ consists of the following three steps:
\begin{enumerate}
\item We interpret the value $p_V(d)$ of the Hilbert polynomial of $V\subseteq\PP^n$ on $d\in\Z$
as the Euler characteristic $\chi(\Oh_V(d))$ of the twisted sheaf $\Oh_V(d)$.
\item The Hirzebruch-Riemann-Roch Theorem~\cite{hirz:66} gives an explicit combinatorial
description of  $\chi(\Oh_V(d))$ in terms of certain determinants $\Delta_\lambda(c)$
(related to Schur polynomials)
in the Chern classes~$c_i$ of the tangent bundle of~$V$.
\item The homology class corresponding to the cohomology class $\Delta_\lambda(c)$ can be
realized up to sign by a degeneracy locus, which is defined as the pullback of a
Schubert variety under the Gauss map (cf.~Fulton~\cite[Ex. 14.3.3]{fult:98}).
We call the geometric degree of such a degeneracy locus a projective character.
\end{enumerate}

This allows to express (certain integer multiples of) the coefficients
of the Hilbert polynomial as integer linear combinations of projective characters.
We now use the fact that the computation of the geometric degree
of varieties is possible in the complexity class $\gGapC$,
and that the class $\gGapC$ is closed under exponential summation (Lemma~\ref{le:gaddition}).
Here $\gGapC$ is a class of functions slightly larger than $\CPCi$, which is
closed under ``generic parsimonious reductions''~\cite{bucl:04}.

\medskip
\noindent{\bf Organization of the article.} In \S\ref{se:prel} we present all the necessary definitions
and facts needed in order to state a formula for the coefficients of the Hilbert polynomial in terms
of projective characters. While the formula is given in \S\ref{se:prel},
the proof is postponed to \S\ref{se:hidegen}.
Section~\ref{se:count} contains background from (counting) complexity theory over $\C$.
In \S\ref{se:compl-hilb} we present the main
results of this article, the upper and lower bounds on the complexity of computing the coefficients of the
Hilbert polynomial.
Finally, \S\ref{se:hidegen} contains the derivation of the relationship between the Hilbert polynomial
and degeneracy loci, using the Hirzebruch-Riemann-Roch theorem. In order to facilitate reading,
the proofs of two technical lemmas from \S\ref{se:prel} and of a result used
in \S\ref{se:compl-hilb} are postponed to the appendix.

\medskip
\noindent{\bf Acknowledgment.} We thank Felipe Cucker for discussions and inviting us to
Hong Kong in Spring~2004, where the basis of this work was elaborated in
the joint papers~\cite{bucl:04a,bucl:04}.

\section{Preliminaries from algebraic geometry}
\label{se:prel}

Throughout this article, unless otherwise stated, the term {\em variety} will mean a complex, projective,
not necessarily irreducible variety. By a {\em subvariety} we will always understand a closed subvariety.
We will say that a property holds {\em for almost all points in
a variety}, if the set of points satisfying the given property is a dense subset
with respect to the Zariski topology.

\subsection{The Hilbert polynomial}
Let $S:=\C[X_0,\dots,X_n]$ denote a polynomial ring and let $M$ be a finitely generated, graded $S$-module. 
Denote by $M_k$ the $k$-th graded part of $M$. The function $h_M\colon \Z\rightarrow \N$, 
defined by $h_M(k)=\dim_{\C}M_k$ is called the {\em Hilbert function} of $M$.
A proof of the following theorem can be found in~\cite[I.7]{hart:77}.

\begin{theorem}[Hilbert-Serre]
Let $M$ be a finitely generated, graded $S$-module. Then there exists a unique polynomial $p_M(T)\in \Q[T]$
such that $h_M(\ell)=p_M(\ell)$ for sufficiently large $\ell$. Furthermore, the degree of $p_M$ equals 
the dimension of the projective zero set
of the annihilator $\{s\in S \ | \ sM=0\}$ of $M$.
\end{theorem}

The polynomial $p_M(T)$ is called the {\em Hilbert polynomial} of $M$. Of special interest is the case
$M=S/I$, where $I\subseteq S$ is a homogeneous ideal. 
If $I=I(V)$ is the homogeneous ideal of a complex projective variety $V\subseteq \PP^n$,
then we write $p_V:=p_{S/I}$ and call this the Hilbert polynomial of $V$. We thus have $\deg p_V=\dim V$.

\begin{example}
\begin{enumerate}
\item The Hilbert polynomial of $V=\PP^n$ is $p_V(T)=\binom{T+n}{n}$.
\item Let $f\in \C[X_0,\dots,X_n]$ be homogeneous and irreducible of degree $d$ and let $V=\mcal{Z}(f)$
be its projective zero set. Then
$p_V(T)=\binom{T+n}{n}-\binom{T+n-d}{n}$. 
\end{enumerate}
\end{example}

Let $V\subseteq \PP^n$ be an $m$-dimensional projective variety with Hilbert polynomial
$p_V(T)=p_mT^m+\dots +p_1T+p_0$. The {\em geometric degree} $\deg V$ of $V$ is defined as
$\deg V:=m! \ p_m$. The degree counts the number of intersection points of $V$ with a generic linear subspace of complementary
dimension \cite[Lect.~18]{harr:92}. It is additive on the irreducible components of maximal dimension.
The {\em arithmetic genus} of $V$ is defined as $g_a(V):=(-1)^m(p_0-1)$. While the degree depends on the embedding in
projective space, the arithmetic genus is a birational invariant (cf.~\cite[Ex.~III.5.3]{hart:77}).

\subsection{Projective characters}
\label{se:proj-char}

General references for the material presented in this section are \cite{fult:97,mani:01}.
In the following we assume $0\le m\le n$. The {\em Grassmann variety}
$$
 \Gr(m,n):=\{A \mid \mbox{$A\subseteq\PP^n$ linear subspace of dimension~$m$}\}
$$
is an irreducible smooth projective variety of dimension $(m+1)(n-m)$ \cite[Lect.~6]{harr:92}.

The {\em flag variety} $\Flag$ is defined as the set of all complete flags
$\uF$ of linear subspaces $F_0\subset \ldots \subset F_{n-1} \subset F_{n}=\PP^n$,
such that $\dim F_i = i$ for $0\le i \le n$.
It is an irreducible smooth projective variety \cite[III.9.1]{fult:97}.

For $A\in\Gr(m,n)$ and a flag $\uF\in\Flag$ we consider the weakly 
increasing sequence of dimensions $(\dim (A\cap F_j))_{0\le j\le n}$ and denote by 
$0\le\sigma_0<\sigma_1<\dots <\sigma_m\leq n$ the positions where 
the ``jumps'' occur, that is, 
$\dim (A\cap F_{j})=i$  for $\sigma_i\leq j <\sigma_{i+1}$ 
(using the conventions $\dim\emptyset = -1$ and $\sigma_{-1} :=0, \sigma_{m+1}:=n$).
The sequence $(\sigma_i)$ can be encoded by the sequence of integers 
$n-m\ge\lambda_1\ge\lambda_2\ge\ldots\ge\lambda_{m+1}\ge 0$ 
defined by $\lambda_{i+1}:=n-m+i-\sigma_i$. 

Generally, a {\em partition} $\lambda=(\lambda_1,\ldots,\lambda_r)$ is a weakly decreasing sequence
of natural numbers.
The length of $\lambda$ is defined as the number of nonzero components of $\lambda$.
The size of~$\lambda$ is defined as $|\lambda|:=\lambda_1 +\cdots+\lambda_r$, and we call 
$\lambda$ {\em a partition of } $k$, if $|\lambda|=k$.
We say that a partition~$\mu$ contains a partition~$\lambda$, $\lambda\subseteq\mu$, if
$\lambda_i \le \mu_i$ for all~$i$ (we set $\lambda_i=0$ for all $i$ exceeding the length of $\lambda$).

To a partition $\lambda$ of length at most $m+1$ with $\lambda_1\le n-m$ 
(in which case we call $\lambda$ {\it admissible}) we associate a strictly increasing sequence
$0\le\sigma_0<\dots <\sigma_m\leq n$ by setting $\sigma_i:=n-m+i-\lambda_{i+1}$
for $0\leq i \leq m$. The~$\sigma_i$ are used to select a subflag
$F_{\sigma_0}\subset \ldots \subset F_{\sigma_{m}}$
with $\dim F_{\sigma_i}=\sigma_i$.
For such a partition~$\lambda$ and a flag $\uF\in\Flag$ the
{\em Schubert variety} $\Omega_\lambda(\uF)$ is defined as follows:
$$
 \Omega_\lambda(\uF) := \{A\in \Gr(m,n) \mid \dim (A\cap F_{\sigma_i}) \ge i\mbox{ for $0\le i \leq m$}\}.
$$
For $A\in\Gr(m,n)$ we always have $\dim (A\cap F_{\sigma_i}) \ge i-\lambda_{i+1}$,
so that $\lambda_{i+1}$ measures the excess in dimension of the intersection.
It is known that $\Omega_\lambda(\uF)$ is an irreducible variety of
codimension~$|\lambda|$ in $\Gr(m,n)$ \cite[III.9.4]{fult:97}. 
(Note that since $\lambda$ is admissible, we have $|\lambda|\leq \dim \Gr(m,n)$.)
In general, Schubert varieties are singular \cite[\S 3.4]{mani:01}.

For a flag $\uF\in\Flag$ and an admissible partition $\lambda$ the
{\em Schubert cell} $e_\lambda(\uF)$ is defined as follows (put $F_{-1}=\emptyset$) 
\begin{equation}\label{eq:def-scell}
 e_\lambda(\uF) := \{A\in \Omega_\lambda(\uF) \mid \dim (A\cap F_{\sigma_{i}-1})=i-1 \mbox{ for $0\leq i\leq  m$}\}.
\end{equation}
Thus $e_\lambda(\uF)$ consists of those elements $A\in \Omega_\lambda(\uF)$ for which $\dim A\cap F_j$ increases
at exactly the positions $j=\sigma_i$.
The Grassmann variety $\Gr(m,n)$ is the disjoint union of the Schubert cells $e_\lambda(\uF)$
over all admissible partitions~$\lambda$. Moreover, it is known that 
\begin{equation}\label{eq:Sdecomp}
 \Omega_\lambda(\uF) = \bigcup_{\lambda\subseteq\mu} e_\mu(\uF),
\end{equation}
where the union is over all admissible partitions~$\mu$ containing $\lambda$, 
cf.~\cite[III.9.4, Ex.~13]{fult:97} or \cite[\S 3.2]{mani:01}. 
The Schubert cell is a complex analytic submanifold of $\Gr(m,n)$ 
of codimension $|\lambda|$. It is open and dense in $\Omega_\lambda(\uF)$.
Moreover, $e_\lambda(\uF)$ is contained in the smooth part of $\Omega_\lambda(\uF)$, cf.~\cite[\S 3.4]{mani:01}.

\begin{example}
\begin{itemize}
\item[(i)]
In the case $\lambda=(k)=(k,0,\ldots,0)$ the degeneracy conditions reduce
to the single condition $A\cap F_{\sigma_0}\ne\emptyset$ on $F_{\sigma_0}\in\Gr(n-m-k,n)$.
\item[(ii)] In the case $\lambda=(1^k)=(1,\ldots,1,0,\ldots,0)$ the degeneracy conditions reduce to
the single condition $\dim(A\cap F_{\sigma_{k-1}}) \ge k-1$ on $F_{\sigma_{k-1}}\in\Gr(n-m+k-2,n)$.

\item[(iii)]
We have $\PP^n=\Gr(0,n)=\Omega_0(\uF)=\cup_{i=0}^ne_{(i)}$, 
where $e_{(i)}=F_i-F_{i-1}\cong \C^i$, which is just the usual decomposition 
of $\PP^n$ as a disjoint union of affine spaces.
\end{itemize}
\end{example}

Let $V\subseteq \PP^n$ be a smooth projective variety of pure dimension~$m$.
The {\em Gauss map} $\varphi\colon V\to~\Gr(m,n)$
maps $x\in V$ to the projective tangent space $\bT_xV\subseteq\PP^n$
at~$x$. For an admissible partition~$\lambda$ and a flag $\uF\in \Flag$ we define
the {\em generalized polar variety}
\begin{equation}\label{eq:def-polar}
 P_\lambda(\uF):=\varphi^{-1}(\Omega_\lambda(\uF))
  =\{x\in V \mid \dim (\bT_x V\cap F_{\sigma_i}) \ge i\mbox{ for $0\le i \le m$}\}
\end{equation}
to be the preimage of the Schubert variety $\Omega_\lambda(\uF)$ under the Gauss map.
The well-known {\em polar varieties}
$$
 P_k(\uF):=P_{(1^k)}(\uF)= \{x\in V \mid \dim (\bT_x V\cap F_{n-m+k-2}) \ge k-1 \}
$$
correspond to the special case $\lambda=(1^k)=(1,\ldots,1,0,\ldots,0)$, see \cite{pien:78,bras:00}.
We remark that a different concept of generalized polar varieties has been previously used for algorithmic purposes,
see~\cite{bghm:01,bghp:04}.

Note that the case where $V$ is a linear space is degenerate:
then $\dim \varphi(V)=0$ and thus $P_\lambda(\uF)$ is empty for almost all $\uF\in\Flag$, provided $|\lambda|>0$.
A result by Zak, cf.~\cite[\S 7]{fula:81}, states that this is the only
degenerate case. Namely, if $V\subseteq\PP^n$ is a nonlinear irreducible smooth projective variety,
then the Gauss map $\varphi\colon V\to\varphi(V)$ is finite. 
In particular, we have $\dim\varphi(V)=\dim V$ in this case.

We recall now the important notion of transversality.
For $x\in V$
we denote by $T_xV$ the Zariski tangent space and by
$d_x\varphi\colon T_xV\to T_{\varphi(x)}\Gr(m,n)$ the
differential of~$\varphi$ at~$x$, respectively.
The Gauss map $\varphi$ {\em meets} the Schubert cell~$e_\lambda(\uF)$
{\em transversely at} $x\in\varphi^{-1}(e_\lambda(\uF))$,
written $\varphi\pitchfork_x e_\lambda(\uF)$, if
$$
 T_{\varphi(x)}\Gr(m,n) = d_x\varphi (T_xV)+T_{\varphi(x)}e_\lambda(\uF) .
$$
Moreover, {\em $\varphi$ meets $e_\lambda(\uF)$ transversely},
written $\varphi\pitchfork e_\lambda(\uF)$, if
$\varphi\pitchfork_x e_\lambda(\uF)$ holds for all $x$ in $\varphi^{-1}(e_\lambda(\uF))$. 

\begin{remark}\label{re:trans}
If $\varphi\pitchfork e_\lambda(\uF)$ then it is well known that 
$\varphi^{-1}(e_\lambda(\uF))$ is a smooth complex submanifold of 
codimension $|\lambda|$ in $V$, unless it is empty. 
(Recall that $e_\lambda(\uF)$ has the codimension $|\lambda|$ in $\Gr(m,n)$.)
\end{remark}

We can extend the notion of transversality to
Schubert varieties in the following natural way, exploiting their stratification (\ref{eq:Sdecomp})
by Schubert cells.

\begin{definition}\label{eq:def-trans}
We say that {\em $\varphi$ meets $\Omega_\lambda(\uF)$ transversely}, written $\varphi\pitchfork \Omega_\lambda(\uF)$, 
if for every admissible $\mu\supseteq \lambda$ we have $\varphi\pitchfork e_\mu(\uF)$.
\end{definition}

The following lemma is proved in Appendix~\ref{se:app-A1}. 

\begin{lemma}\label{le:proj-char}
Let $V\subseteq\proj^n$ be a smooth projective variety of pure dimension $m$ such that not
all irreducible components of $V$ are linear.
Let $\varphi\colon V\to\Gr(m,n)$ be the Gauss map of $V$
and $\lambda$ be an admissible partition with $|\lambda|\leq m$.
Then we have
\begin{description}
\item[(i)]
$\varphi\pitchfork \Omega_\lambda(\uF)$ for almost all flags $\uF \in\Flag$,
\item[(ii)]
if $\varphi\pitchfork \Omega_\lambda(\uF)$, then 
$\dim (\varphi(V)\cap e_\lambda(\uF)) = m -|\lambda|$ and 
$\codim_V P_\lambda(\uF) = |\lambda|$, 
\item[(iii)]
there exists an integer $d_\lambda$, such that $\deg P_\lambda(\uF)=d_\lambda$,
provided $\varphi\pitchfork \Omega_\lambda(\uF)$.
\end{description}
\end{lemma}

We call $\deg P_\lambda:=d_\lambda$ the {\em projective character} of $V$ corresponding to~$\lambda$.
These quantities were studied by Severi~\cite{seve:02}, see also~\cite[Ex.~14.3.3]{fult:98}.
Note that the degree of~$V$ equals the projective character for $\lambda=0$.

\begin{example}\label{ex:curve}
Let $V\subseteq \PP^2$ be a smooth curve. Then $\deg P_1$ counts the number of points on the
curve whose tangents go through a generic point in $\PP^2$. B\'ezout's theorem implies
that this number equals $d(d-1)$, where $d$ is the degree of the curve.
\end{example}

The following will be used later. Again, the proof is postponed to the Appendix.

\begin{lemma}\label{le:trans-Schubert}
Let $W$ be a quasiprojective variety and let $\psi\colon W\to\Gr(m,n)$ be a morphism.
Let $\lambda$ be an admissible partition. 
For $\uF\in\Flag$ set $R_\lambda(\uF):=\psi^{-1}(\Omega_\lambda(\uF))$.
Then for almost all $\uF\in\Flag$ we have
$\dim R_\lambda(\uF) \le \dim W -|\lambda|$
if $|\lambda|\le\dim W$, and $R_\lambda(\uF)=\emptyset$ otherwise.
\end{lemma}

\subsection{Expressing the Hilbert polynomial by projective characters}

Our goal is to express the coefficients of the Hilbert polynomial of $V$ in terms of its
projective characters.
We first introduce some notation.

To any sequence $c=(c_i)_{i\in \N}$ of elements of a commutative ring such that $c_0=1$
and to a partition $\lambda=(\lambda_1,\ldots,\lambda_r)$ we assign the ring element $\Delta_\lambda(c)$
as follows:
\begin{align}\label{eq:defDelta}
  \Delta_{\lambda}(c) & :=\det\left((c_{\lambda_i-i+j})_{1\leq i,j\leq r}\right)\notag \\
  & =\det \begin{pmatrix}
                         c_{\lambda_1} & c_{\lambda_1+1} & \cdots & c_{\lambda_1+r-1}\\
                         c_{\lambda_2-1} & c_{\lambda_2} & \cdots & c_{\lambda_2+r-2}\\
                         \cdots & \cdots & \cdots & \cdots\\
                         c_{\lambda_r-r+1} & c_{\lambda_r-r+2} & \cdots & c_{\lambda_r}
                       \end{pmatrix},
\end{align}
using the convention $c_i=0$ for $i<0$. Note that the value of this determinant does not change
if we extend the partition $\lambda$ by zeros.

In the following let $b$ be the coefficient sequence of the power series
\begin{equation}\label{eq:def-b}
\sum_{i\geq 0}b_i t^i := \frac{t}{1-e^{-t}} = 1 + \frac{t}{2} + \sum_{j\ge 1} (-1)^{j-1} \frac{B_j}{(2j)!}\,t^{2j},
\end{equation}
where the $B_j$ are the {\em Bernoulli numbers}. E.g.,
$B_1=\frac{1}{6},B_2=\frac{1}{30},B_3=\frac{1}{42}$.

\begin{remark}\label{re:integer}
It is known that
$B_n = (-1)^{n-1} \sum_{k=1}^{2n} \frac{1}{k+1}\sum_{r=1}^k (-1)^r \binom{k}{r} r^n$ \cite{mwbe:04}.
This implies that $(2n+1)! B_n$ is an integer, hence
$i!(i+1)! b_i$ is an integer for all $i$.
Taking into account that for a partition $\lambda=(\lambda_1,\ldots,\lambda_r)$
of size~$M$ and length $r$ we always have $\lambda_1 + r -1 \le M$, we conclude that
$[(M+1)!\cdots(M-r+2)!]^2\,\Delta_\lambda(b)$ is an integer.
\end{remark}

To a pair $(\lambda,\mu)$ of partitions of length at most~$m$ we assign the following determinant of
binomial coefficients
\begin{equation*}
 d^m_{\lambda \mu}:=\det \left(\binom{\lambda_i+m+1-i}{\mu_j+m+1-j}\right)_{1\leq i,j\leq m}.
\end{equation*}
Now let $0\le k\le m$ and $\mu$ be a partition with $|\mu|\leq m-k$.
To this data we assign the rational number
\begin{equation}\label{eq:defdelta}
 \delta^{m,k}_\mu := (-1)^{|\mu|}\sum_{\substack{\mu\subseteq \lambda\\ |\lambda|=m-k}}
                        \Delta_{\lambda}(b)d^m_{\lambda \mu},
\end{equation}
where the sum is over all partitions $\lambda$ of size $m-k$ that contain $\mu$ as subpartition.

The following crucial statement will be proved in \S\ref{se:hidegen}.

\begin{theorem}\label{th:hmain}
Let $V\subseteq \PP^n$ be a smooth complex projective variety of pure dimension~$m$ and $0\le k\le m$.
Then the $k$-th coefficient $p_k(V)$ of the Hilbert polynomial of $V$
is given by
\begin{equation*}
  p_k(V)= \frac{1}{k!}\sum_{\substack{|\mu|\leq m-k\\ \mu_1\leq n-m}}\delta^{m,k}_{\mu}\deg P_{\mu},
\end{equation*}
where $\deg P_{\mu}$ is the projective character introduced in \S\ref{se:proj-char}.
In particular, $[(m-k+1)!\cdots 2! 1!]^2\, k!\,p_k(V)$ is an integer.
\end{theorem}

\begin{example}
\begin{enumerate}
\item The above formula yields
$p_m(V) = \frac{1}{m!}\,\delta^{m,m}_{0}\deg P_{0} = \frac{1}{m!}\,\deg V$,
as expected (check that $\Delta_0(b)=1,d^m_{0,0}=1$).

\item In the case where $V\subseteq\PP^n$ is a smooth curve ($n\geq 2$), the above formula implies that
$p_0= \delta^{1,0}_0 \deg P_0 + \delta^{1,0}_1 \deg P_{1}= \deg V - \frac{1}{2}\deg P_{1}$,
where $\deg P_1 = \#\{x\in V \mid \bT_xV \cap L \neq\emptyset \}$
for a generic linear subspace $L\subset\PP^n$ of codimension~$2$.

\item In the special case of a smooth planar curve~$V$ (see Example \ref{ex:curve}), we have
$p_0(V)= d - \frac{1}{2}d(d-1) = \frac{1}{2}d(3-d)$, which implies the well known formula
$1-p_0(V) = \frac{1}{2}(d-1)(d-2)$ for the arithmetic genus.

\item Consider the rational normal curve $V\subseteq\PP^n$, which is defined as the projective
closure of $\{(t,t^2,\ldots,t^n) \mid t\in\C\}$. The Hilbert polynomial of $V$
satisfies $p_V(T) = n T + 1$. It is not too hard to verify directly that
$\deg P_1 = 2(n-1)$.

\end{enumerate}
\end{example}

\section{Counting complexity over the complex numbers}
\label{se:count}

We will consider BSS-machines over $\C$ as they are defined
in~\cite{blss:89,bcss:95}. Roughly speaking, such a
machine takes an input from $\Ci$, performs a
number of arithmetic operations and tests for zero
following a finite list of instructions, and
halts returning an element in $\Ci$ (or loops forever).
The computation of a machine on an input $x\in\Ci$
is well-defined and notions such as a function being
computed by a machine or a subset of $\Ci$ being decided
by a machine easily follow.
A machine $M$ over $\C$ is said to work in polynomial time if
there is a constant $c\in \N$ such that for every
input $x\in \Ci$, $M$ reaches its output node after at
most $\size(x)^c$ steps.
Hereby, we define $\size(x)$
to be the smallest $n\geq 0$ such that $x\in \C^n$.
The complexity classes $\PC$ and $\NPC$ are defined as usual, as well as the
notions of reduction and completeness.
The class $\coNPC$ consists of all subsets of $\Ci$ whose complement lies in $\NPC$.
In a completely analogous fashion we can also consider machines over $\R$ and
corresponding classes $\PR$ and $\NPR$.

In~\cite{blss:89} it was shown that the following fundamental problems is $\NPC$-complete.
\begin{description}
\item{$\HNC$} (Hilbert's Nullstellensatz) \quad
Given a finite set of complex multivariate polynomials, decide whether these polynomials have a common zero.
\end{description}
For our convention on coding polynomials as elements of $\Ci$ we refer to \S\ref{se:compl-hilb},
see also \cite[1.2]{koir:97-1} for a discussion.

\subsection{Counting complexity classes}

In classical complexity theory, Valiant~\cite{vali:79-1} introduced the counting class
$\#\P$ as the class of functions which count the number of accepting paths of nondeterministic polynomial
time Turing machines. One of his main results~\cite{vali:79-2}
was that the problem of counting the number of perfect matchings in a bipartite
graph, or equivalently, computing the permanent of its adjacency matrix, is $\#\P$-complete.
For a comprehensive account to counting complexity
we refer to \cite[Chapter 18]{papa:94} and \cite{fort:97}.

We now recall the definition of counting classes over $\C$
from~\cite{bucu:04,bucl:04}, which follows the lines used in
discrete complexity theory to define $\#\P$ and $\GapP$~\cite{fort:97}.
We denote by $\eZ:=\Z\cup\{-\infty,\infty,\mathrm{nil}\}$ the union of the set $\Z$ with three additional symbols
$-\infty,\infty$, and $\mathrm{nil}$ (the latter standing for undefined). For the arithmetic in $\eZ$, we refer to \cite[\S 4.3]{bucl:04}.

\begin{definition}\label{def:sp}
\begin{description}
\item[(i)] A function $\varphi\colon\Ci\to\N\cup\{\infty\}$ belongs
to the class $\CPCi$ if there exists a
polynomial time machine~$M$ and a polynomial $p$ such that
$\varphi(x)=|\{y\in\C^{p(n)}\mid M\mbox{ accepts } (x,y)\}|$
holds for all $x\in\C^n$ and all $n\in\N$.

\item[(ii)] The class $\GapC$ consists of all functions
$\gamma\colon\Ci\to\eZ$ of the form
$\gamma=\varphi-\psi$ for $\varphi,\psi\in\CPCi$.
\end{description}
\end{definition}

We next define notions of reduction and completeness for counting classes.

\begin{definition}\label{def:reduction}
\begin{enumerate}
\item Let $\varphi,\psi\colon\Ci\to\eZ$. We say that
$\pi\colon\Ci\to\Ci$ is a {\em parsimonious reduction} from
$\varphi$ to $\psi$ if $\pi$ can be computed in polynomial
time and, for all $x\in\Ci$, $\varphi(x)=\psi(\pi(x))$.

\item Let $\varphi,\psi\colon\Ci\to\Ci$. We say that $\varphi$ {\em Turing reduces to $\psi$}
if there exists an oracle machine which, with oracle $\psi$, computes $\varphi$ in polynomial time.
\end{enumerate}
\end{definition}

Let $\cal C$ be either $\CPCi$, or $\GapC$.
We say that a function $\psi$ is {\em hard} for $\cal C$
if for every $\varphi\in{\cal C}$ there is a parsimonious
reduction from $\varphi$ to~$\psi$. We say that $\psi$ is {\em
$\cal C$-complete} if in addition $\psi\in{\cal C}$.
The notions of {\em Turing-hardness} and {\em Turing-completeness}
are defined similarly.

In~\cite{bucu:04} it was shown that the following basic problem
is  $\CPCi$-complete with respect to parsimonious reductions.
\begin{description}
\item{$\#\HNC$} ({\em Algebraic point counting})\quad
Given a finite set of complex multivariate polynomials,
count the number of complex common zeros,
returning $\infty$ if this number is not finite.
\end{description}

In an analogous fashion, the problem $\GapHNC$ (as introduced in \cite[\S 4.3]{bucl:04}), which
counts the difference in the the number of solutions of two given systems of polynomial equations, 
is seen to be complete for the class $\GapC$.

There are algorithms solving $\#\HNC$ in single exponential time
(or even parallel polynomial time).
A key point for showing this is the fact that a Gr\"obner basis of a zero-dimensional ideal can be
computed in single exponential time \cite{difg:91,laks:90,lala:90}.
The number of solutions can then be determined using linear algebra techniques,
as described for example in \cite[Chapter 2]{tapas:99}.

We remark that a corresponding counting class $\CPRi$ over the reals has been introduced by
Meer \cite{meer:00} and was further explored in \cite{bucu:04}.

\subsection{Polynomial hierarchy over the reals}

The constant-free polynomial hierarchy over the reals will be needed in the next section
for extending the notion of a parsimonious reduction.
It is important to work over the reals since the polynomial hierarchy over the complex numbers
has not enough expressive power for our purposes.

For what follows, we call a machine (over $\R$ or $\C$) {\em constant-free},
if its only machine constants are $0$ and $1$.
The following definition is from~\cite[Chapter 21]{bcss:95}.

\begin{definition}\label{def:ph}
A relation $R\subseteq \Ri$ is said to be in $\Sigma_k^0$ for $k\in \N$,
if there exists a relation $A\subseteq (\Ri)^{k+1}$,
decidable in polynomial time by a constant-free machine $M$ over $\R$,
and polynomials $p_1,\dots,p_k$, such that for $x\in \R^n$:
\begin{equation*}
  x\in R \Leftrightarrow Q_1x_1\in \R^{p_1(n)}\dots Q_kx_k\in \R^{p_k(n)}\ (x_1,\dots,x_k,x)\in A
\end{equation*}
where $Q_1=\exists$ and the quantifiers $Q_i\in \{\exists,\forall\}$ alternate.
We define the {\em constant-free polynomial hierarchy} $\PHR^0$ to be
the union $\PHR^0=\cup_k \Sigma_k^0$.
\end{definition}

We next observe that the dimension and local dimension of semialgebraic sets
can be expressed in $\PHR^0$.
We study this in the general situation of a family of semialgebraic sets~$S_u$
depending on a parameter $u\in\Ri$ such that the property
$x\in S_u$ is expressible in $\PHR^0$.
We denote by $\dim_x S_u$ the local dimension of $S_u$ at~$x\in S_u$
(defined to be $-1$ if $x\not\in S_u$).

\begin{lemma}\label{le:dim-ph}
Let $R\subseteq\Ri\times \Ri$ be a relation in $\PHR^0$, $p$~be a polynomial, and
consider for $u\in\R^{n}$ the semialgebraic set
$S_u:=\{x\in\R^{p(n)}\mid (u,x)\in R\}$.
Then both decision problems $\{(u,d)\in\Ri\times\N\mid\dim S_u\geq d\}$
and $\{(u,x,d)\in\Ri\times\Ri\times\N\mid\dim_x S_u\geq d\}$ are in $\PHR^0$.
\end{lemma}

\begin{proof}
We have $\dim S_u\ge d$ if and only if there exists a $d$-dimensional coordinate subspace
such that the projection of $S_u$ on this subspace has a nonempty interior.
Writing this condition as a first order formula over $\R$ yields the claim for
the dimension.
(For a more economic description, see~\cite{koir:99a}.)

Let $B_\epsilon(x)$ denote the open ball with radius~$\epsilon$ centered at~$x$.
We have $\dim_x S_u \ge d$ if and only if $\dim (S_u\cap B_\epsilon(x)) \ge d$ for
sufficiently small $\epsilon>0$, cf.~\cite{bapr:03}.
Writing this as a first order formula over $\R$ implies the claim about the
local dimension.
\end{proof}

\subsection{Generic parsimonious reductions}

The concept of generic parsimonious reduction, as introduced in \cite{bucl:04}
and implicit in \cite{bucu:04},
allows to make ``general position'' arguments as part of a reduction algorithm.
A paradigmatic example is that of reducing the problem of computing the geometric degree
of a variety $V$ to $\HNC$ by intersecting~$V$ with a generic linear subspace of complementary
dimension. We are interested in problems where it is possible to compute in polynomial time
a list of candidates for generic parameters, among which the majority is in fact ``generic''
(see the notion of partial witness sequences introduced in~\cite{bucu:04}).
This can be achieved by only requiring the genericity condition to be describable in
terms of the constant-free polynomial hierarchy over the reals.

In the following, we are concerned with relations $R\subseteq \Ci\times \Ci$.
It makes sense to say that such a relation is in $\PHR^0$
by representing points in $\C^n$ as points in $\R^{2n}$ in the obvious way.

We call a relation $R\subseteq\Ci\times\Ci$ {\em balanced} if there is a polynomial~$p$ such that
$R(u,a)$ implies $\size(a)\le p(\size(u))$ for all $(u,a)\in\Ci\times\Ci$. In this case we say
that $p$ is associated to $R$.
Moreover, we will write $\forall^\ast a\in\C^n\, R(a)$ in order to express that
Zariski almost all points $a\in\C^n$ satisfy a relation $R$.

\begin{definition}\label{def:genpr}
Let $\varphi,\psi\colon\Ci\to\hat{\Z}$.
A {\em generic parsimonious reduction} from $\varphi$ to $\psi$
consists of a pair $(\pi,R)$, where $\pi\colon\Ci\times\Ci\to\Ci$ is
computable in polynomial time over $\C$ by a constant-free machine,
and $R\subseteq\Ci\times\Ci$ is a balanced relation (with associated polynomial $p$)
in $\PHR^0$ such that for all $m\in \N$ the following holds:
\begin{description}
\item[(i)] $\forall u\in\C^m\ \forall a\in\C^{p(m)}\
 ( R(u,a)\Rightarrow \varphi(u)=\psi(\pi(u,a)))$,
\item[(ii)] $\forall u\in\C^m\ \forall^\ast a\in\C^{p(m)}\ R(u,a)$.
\end{description}
\end{definition}

We write $\varphi\gpr\psi$ if there exists a generic parsimonious reduction
from $\varphi$ to $\psi$.
In \cite{bucl:04} it was shown that this is a transitive relation.
The following important fact is shown in \cite[Theorem 4.4]{bucl:04}.

\begin{theorem}\label{th:gen-turing}
Let $\varphi,\psi\colon \Ci\rightarrow \eZ$. If $\varphi\gpr\psi$ then $\varphi$ Turing reduces to~$\psi$~.
\end{theorem}

The closures of $\CPCi$ and $\GapC$ with respect to generic parsimonious reductions
defined below seem to capture more accurately the kind of counting problems encountered in algebraic geometry.

\begin{definition}\label{def:genSP}
\begin{description}
\item[(i)] The class $\gCPCi$ consists of all
functions $\varphi\colon\Ci\to\N\cup\{\infty\}$ such that there exists
$\psi\in\CPCi$ with $\varphi\gpr\psi$.
\item[(ii)]
The class $\gGapC$ consists of all functions
$\varphi\colon\Ci\to\hat{\Z}$
such that there exists $\psi\in\GapC$ with $\varphi\gpr\psi$.
\end{description}
\end{definition}
The functions in $\gGapC$ can also be characterized as the differences of two functions in $\gCPCi$. 

Similar as $\GapP$ (cf. \cite{fort:97}), the class $\gGapC$ is closed under exponential summation.

\begin{lemma}\label{le:gaddition}
Let $\varphi\colon\Ci\x\{0,1\}^\infty\to\Z$ be a function in $\gGapC$,
$q$ be a polynomial, and $g\colon \{0,1\}^\infty\to\Z$ be in $\GapP$.
Define $\tilde{\varphi}\colon\Ci\to\Z$ by setting for $u\in\C^m$
\begin{equation*}
  \tilde{\varphi}(u) =\sum_{y\in\{0,1\}^{q(m)}} g(y) \varphi(u,y)
\end{equation*}
Then $\tilde{\varphi}$ belongs to $\gGapC$. A similar statement holds for $\GapC$.
\end{lemma}

\begin{proof}
The function
$\Ci\x\{0,1\}^\infty\to\Z, (u,y)\mapsto g(y)\varphi(u,y)$ is in $\gGapC$
since the product of two finite valued functions in $\gGapC$ is in $\gGapC$~\cite[Lemma 4.9]{bucl:04}.
The claim now follows from the fact that $\gGapC$ is closed under
exponential summation~\cite[Lemma 4.10]{bucl:04}.
\end{proof}

In~\cite{bucu:04} the problem of computing the geometric degree of the zero set $Z\subseteq\C^n$
of given complex polynomials was Turing reduced to $\HNC$.
An analysis of the proof reveals that this reduction is generic parsimonious
except for the computation of the dimension of~$Z$ at the beginning.
Therefore, the following slight modification of the degree problem is in $\gCPCi$: given~$Z$ as above
and $d\in\N$ such that $\dim Z\le d$, compute the geometric degree of the $d$-dimensional part of $Z$.

We can even extend this to the situation, where $Z_u\subseteq\C^n$ is a constructible
set depending on a complex parameter vector~$u$ and membership of $x$ in $Z_u$ can be
decided by a polynomial time machine.
(By the degree of a constructible set we understand the sum of the degrees of its components
of maximal dimension.)

\begin{lemma}\label{le:deg-mach}
Let $M$ be a polynomial time machine over $\C$, $p\colon\N\to\N$ be a polynomial and
consider for $u\in\C^n$ the  constructible set
$$
 Z_u:=\{x\in\C^{p(n)}\mid \mbox{ $M$ accepts $(u,x)$}\} .
$$
Then there is a function $\varphi$ in $\gCPCi$ such that for all $u\in\Ci,d\in\N$
the value~$\varphi(u,d)$ equals the degree of the $d$-dimensional part of $Z_u$,
provided $\dim Z_u\le d$.
\end{lemma}

\begin{proof}
The proof is completely analogous to the case where $Z_u$ is given as the zero set of polynomials
\cite{bucu:04}, see also \cite[Theorem 7.2]{bucu:04b}.
\end{proof}

\begin{example}\label{ex:dim-mach}
Let $F$ be a matrix with entries in $\C[X_1,\dots,X_{n}]$, $k,d\in\N$ such that
$Z:=\{ x\in \C^{n} \mid \mathrm{rank}\, F(x)\leq k\}$ has dimension at most~$d$.
Then, by Lemma~\ref{le:deg-mach}, the degree of $Z$ can be computed in $\gCPCi$.
This follows since the rank condition can be tested
in polynomial time using linear algebra.
(However, writing down the rank condition in terms of non-vanishing of minors
would lead to a representation of exponential size.)
\end{example}

\subsection{Boolean parts}\label{se:boole}

If $L$ is one of the computational problems studied in this paper,
we denote by $L^\Z$ its restriction to input polynomials with integer coefficients.
These discrete problems will be studied in the Turing model of computation.
For doing so, we introduce the notion of the Boolean part of a complexity class over~$\C$.
For more details on the following we refer to~\cite{bucu:04, bucl:04}. 

Let $\mcal{C}$ be a class of functions $\Ci\rightarrow \hat{\Z}$.
We define its {\em Boolean part} $\BP(\mcal{C})$ as the
class of functions $\{0,1\}^{\infty}\rightarrow \eZ$
obtained from functions in $\mcal{C}$ by restriction to $\{0,1\}^{\infty}$.
In~\cite{bucu:04}, the class $\GCC$ of {\em geometric counting complex problems} 
was defined as the Boolean part of $\CPCi$.
The discrete version $\#\HNCbit$ of the problem $\#\HNC$
is $\GCC$-complete~\cite{bucu:04}.
Both $\GCC$ and the Boolean part $\BP(\gGapC)$ are closed under parsimonious reductions and 
(cf.~\cite{bucl:04}) 
\begin{equation*}
  \CP\subseteq\GCC\subseteq \BP(\gGapC) \subseteq \FPSPACE.
\end{equation*}
The class $\BP(\gGapC)$ is closely related to $\GCC$ in the sense that any 
$\varphi\in\BP(\gGapC)$ can be parsimoniously reduced to $\GapHNC$ in 
a randomized sense, cf.~\cite[Remark~6.7]{bucl:04}.  
It is a challenging open problem to characterize $\GCC$ and $\BP(\gGapC)$ 
in terms of previously studied classical complexity classes. 

\section{Complexity of computing the Hilbert polynomial}
\label{se:compl-hilb}

Our goal is to show that the problem of computing the Hilbert polynomial of
a smooth equidimensional projective variety lies in the class $\gGapC$.
When trying to formally define the problem
under consideration, the question arises whether the smoothness condition can
be tested at all within these resources. The obvious idea of checking the
Jacobian criterion at all points in the variety $V$ (which is possible in
$\coNPC$) will fail if the given polynomials $f_1,\ldots,f_r$ describing the variety~$V$
do not generate a radical ideal and thus differ from the vanishing ideal $I(V)$ of $V$.
Indeed, it is not known whether a set of generators of $I(V)$ can be computed from
$f_1,\ldots,f_r$ in parallel polynomial time or even weaker, in single exponential time.

We overcome these difficulties by requiring an input specification, which, on the one hand,
can be checked in $\coNPC$, and on the other hand guarantees
that the highest dimensional part of the variety is smooth. The goal is then to compute the
Hilbert polynomial of the highest dimensional part.

Thus in the following, we will assume that the projective variety
$V'\subseteq\PP^n$ is given as the zero set of a family
$f_1,\ldots,f_r$ of homogeneous polynomials in $\C[X_0,\dots,X_n]$
satisfying the following {\em input condition}:
\begin{equation}\label{eq:input-cond}
\forall x\in \mZ(f_1,\ldots,f_r)-\{0\}\ \dim \{z\in\C^{n+1}\mid
d_xf_1(z)=0,\ldots,d_xf_r(z)=0\} \le m+1
\end{equation}
for some $m\in\N$.
Here and in the following,
we assume that the given polynomials~$f$ are encoded as strings in $\Ci$ using the {\em sparse encoding}.
Thus $f=\sum_{e\in I}a_eX_0^{e_0}\cdots X_n^{e_n}$
is represented by a list of pairs $(a_e,e)$, where the coefficients $a_e$ are given as complex numbers,
while the exponent vector $e$ is given by a bit vector of length at most $\OO(n\log \deg f)$.

We remark that the input condition~(\ref{eq:input-cond}) can be tested in $\coNPC$.

\begin{lemma}\label{le:input}
Assume that the input condition~(\ref{eq:input-cond}) is satisfied. Then
$V'= V \cup W$ is a disjoint union of a smooth variety $V\subseteq\PP^n$ of pure dimension~$m$
(possibly empty) and a subvariety $W\subseteq\PP^n$ with $\dim W <m$.
In particular, the irreducible components of the $m$-dimensional part~$V$ of $V' $are pairwise disjoint.
A point $x\in V'$ is in $V$ if and only if $\dim_x V' =m$.
Moreover, $n-m\le r$ and for all $x\in V$
the Jacobian matrix $(\frac{\partial f_s}{\partial X_i}(x))$ has rank $n-m$.
\end{lemma}

\begin{proof}
For all $x\in V'$ we have $\bT_xV'\subseteq \proj(\cap_{i=1}^r \ker d_xf_i)$,
where $\bT_xV'$ is the projective tangent space of $V'$ at~$x$.
The input condition implies that
for all $x\in V'$, $\dim_x V'\le \dim\bT_xV' \le m$ holds.
Therefore, all points $x\in V'$ of local dimension~$m$ are smooth.
The first claim follows since there is exactly one irreducible component
passing through a smooth point.
The remaining claims are clear.
\end{proof}

We give now a formal definition of our main problem under investigation.
In order to make sure that the output is an integer, we require to compute
a certain multiple of the $k$-th coefficient of the Hilbert polynomial.

\begin{description}
\item{$\HILBS$} ({\em Hilbert polynomial of smooth equidimensional varieties})\quad
Given integers $0\leq k\leq m\leq n$ and a family $f_1,\ldots,f_r$ 
of homogeneous polynomials in $\C[X_0,\dots,X_n]$
satisfying the input condition for $m$,
compute the integer multiple $N(k,m)\, p_k(V)$ 
of the $k$-th coefficient $p_k(V)$ of the Hilbert polynomial 
of the $m$-dimensional part~$V$ of~$V'$, 
where $N(k,m):=[(m-k+1)!\cdots 2! 1!]^2$.  
\end{description}
Here is the main result of this article.

\begin{theorem}\label{th:hilbs-ub}
The problem $\HILBS$ is in $\gGapC$. In particular, the problem $\HILBS$ Turing reduces 
(over $\C$) to $\HNC$.
\end{theorem}

This theorem immediately implies the following corollary, cf. Section \ref{se:boole}.
Recall that $\HILBSbit$ denotes the restriction of $\HILBS$ to input polynomials with integer coefficients.

\begin{corollary}\label{cor:hilbs-ub}
The problem $\HILBSbit$ is in $\BP(\gGapC)$. 
In particular, the problem $\HILBSbit$ Turing reduces to $\HNCbit$ (in the sense of classical 
Turing machines).
\end{corollary}

\subsection{Upper bounds}

The upper bound on $\HILBS$ is based on Theorem \ref{th:hmain}. We therefore first study the problem
to compute projective characters (recall Lemma \ref{le:proj-char} for their definition).

\begin{description}
\item{$\Projchar$} ({\em Projective characters})\quad
Given $0\leq m\leq n$, homogeneous polynomials $f_1,\ldots,f_r$ in $\C[X_0,\dots,X_n]$
satisfying the input condition for~$m$ and a partition $\lambda$
such that $\lambda_1\le n-m$ and $|\lambda|\le m$, compute the 
projective character $\deg P_\lambda$ of
the $m$-dimensional part $V$ of~$V'=\mcal{Z}(f_1,\dots,f_r)$.
\end{description}

\begin{proposition}\label{pro:projchar}
The problem $\Projchar$ is in $\gCPCi$.
\end{proposition}

Using this proposition, we can immediately proceed to prove the main Theorem~\ref{th:hilbs-ub}.

\proofof{Theorem~\ref{th:hilbs-ub}}
Put $N(k,m):= [(m-k+1)!\cdots 2! 1!]^2$. 
Consider the function $g\colon \{0,1\}^{\infty}\rightarrow \Z$ mapping 
$(m,k,\mu)$ to $N(k,m)\delta_{\mu}^{m,k}$,
where $m,k\in \N$, $\mu$ a partition with $|\mu|\leq m-k$, $\mu_1\leq n-m$ and $\delta_{\mu}^{m,k}$ 
is defined in Equation~(\ref{eq:defdelta}), i.e.,
$\delta^{m,k}_\mu := (-1)^{|\mu|}\sum_{\mu\subseteq \lambda, |\lambda|=m-k}\Delta_{\lambda}(b)d^m_{\lambda \mu}$.
By Remark~\ref{re:integer}, the values of $g$ are integers.
The functions mapping $(m,k,\mu,\lambda)$ to $\Delta_{\lambda}(b)d^m_{\lambda \mu}$ and to $N(k,m)$, 
respectively, are clearly polynomial time computable, if
we think of $(m,k,\mu)$ as being encoded in unary.
It then follows from elementary properties of $\GapP$ (closure under exponential summation and product, cf.~\cite{fort:97})
that $g$ is in $\GapP$.
Let $\varphi\colon \C^{\infty}\times \{0,1\}^{\infty}\rightarrow \Z\cup \{-\infty,\infty\}$ be the function
corresponding to the problem $\Projchar$, where the first argument contains the description of the polynomials and
the second argument the partition $\lambda$. According to Proposition~\ref{pro:projchar}, $\varphi\in \gCPCi$,
so we can apply the Summation Lemma~\ref{le:gaddition} to
the main formula in Theorem~\ref{th:hmain} to conclude that $\HILBS\in \gGapC$.
\proofend

We prove Proposition \ref{pro:projchar} using a generic parsimonious reduction from $\Projchar$ to
a certain auxiliary problem, which we describe next.
Consider an instance of $\Projchar$. Write
$\psi(x):=\mathbb{P}\big(\bigcap_{i=1}^r \ker d_xf_i\big)$ for $x\in V'$ and define
for a flag $\uF\in\Flag$ the following constructible set (recall that $\sigma_i = n-m+i-\lambda_{i+1}$)
\begin{equation}\label{Qdef}
 Q_\lambda(\uF):=\{x\in V'\mid \dim (\psi(x)\cap F_{\sigma_i}) \ge i\mbox{ for $0\le i \le m$}\}.
\end{equation}
We will represent a flag $\uF\in \Flag$ by a matrix $a\in\C^{n\times (n+1)}$
such that $F_{\sigma_i}$ is the projective zero set of the linear forms corresponding to the first
$\delta_i:= n -\sigma_i = m - i +\lambda_{i+1}$ rows of $a$, for $0\leq i<m$.

\begin{lemma}\label{le:aux}
There is a function $\Phi$ in $\gCPCi$ which takes as input an instance of $\Projchar$ and a flag $\uF\in \Flag$
and outputs the degree of the $(m-|\lambda|)$-dimensional part of $Q_\lambda(\uF)$,
provided $\dim Q_\lambda(\uF)\le m-|\lambda|$.
\end{lemma}

\begin{proof}
Suppose we have an instance of $\Projchar$ and a flag $\uF\in\Flag$ given by the matrix
$a\in \C^{n\times(n+1)}$.
Let $M_i(x,a)\in\C^{(\delta_i+r)\times (n+1)}$ denote the matrix
obtained by taking the submatrix of $a$ consisting of the first $\delta_i$
rows of~$a$ and adding the Jacobian matrix $(\partial f_s/\partial X_j(x))_{1\le s\le r, 0\le j\le n}$
at the bottom. Then we have for all $x$
$$
 \dim(\psi(x)\cap F_{\sigma_i}) \ge i \Longleftrightarrow \mathrm{rank} M_i(x,a) \le n-i .
$$
This condition can be tested in $\PC$, since the rank of a matrix can be computed in polynomial time,
e.g., using Gaussian elimination (compare Example~\ref{ex:dim-mach}).
The claim follows now from Lemma~\ref{le:deg-mach}.
\end{proof}

\Proofof{Proposition \ref{pro:projchar}}
Suppose we are given an instance of $\Projchar$.
Let $\psi(x)=\mathbb{P}\big(\bigcap_{i=1}^r \ker d_xf_i\big)$
and $Q_\lambda(\uF)$ be defined for a flag $\uF\in \Flag$ as in~(\ref{Qdef}).
By the input condition~(\ref{eq:input-cond}), $\psi(x)$ is a linear subspace of $\PP^n$ of dimension at most~$m$
for every $x\in V'$. Let $V'=V\cup W$ be as in Lemma~\ref{le:input}, so that $V$ is smooth of dimension $m$ and $\dim W<m$.
We then have $\psi(x)=\bT_xV$ for all $x\in V$, so that the restriction $\varphi:=\psi|_V$
determines the Gauss map $\varphi\colon V\rightarrow \Gr(m,n)$. Note that $\psi(x)$ may be different from the projective tangent space
at points $x\in W$.

Set $P_\lambda(\uF):=Q_\lambda(\uF)\cap V$ and $R_\lambda(\uF):=Q_\lambda(\uF)\cap W$.
Then $P_\lambda(\uF)$ is the generalized polar variety introduced in (\ref{eq:def-polar})
and we have $Q_\lambda(\uF)=P_\lambda(\uF)\cup R_\lambda(\uF)$.

Consider the following property of an instance~$I$ of $\Projchar$ and a flag $\uF\in \Flag$:
\begin{equation}\label{eq:cond}
\mbox{ $\varphi \pitchfork \Omega_\lambda(\uF)$ and $\dim R_\lambda(\uF)<m-|\lambda|$. \tag{$\Pi$}}
\end{equation}
According to Lemma \ref{le:proj-char}, the condition $\varphi \pitchfork \Omega_\lambda(\uF)$ 
implies that $\dim P_\lambda(\uF) = m -|\lambda|$ and $\deg P_\lambda(\uF) = \deg P_\lambda$, 
under the assumption that not all components of $V$ are linear, or $\lambda=0$. 
(If the latter assumption is violated, then $P_\lambda(\uF)=\emptyset$.) 
We therefore get
\begin{equation*}
\Pi \text{ is satisfied } \Longrightarrow  \deg P_\lambda = \deg P_\lambda(\uF) = \Phi(I,\uF),
\end{equation*}
where $\Phi$ is the function from Lemma~\ref{le:aux}, i.e., the degree of the $(m-|\lambda|)$-dimensional part of $Q_\lambda(\uF)$.
This establishes a generic parsimonious reduction from $\Projchar$ to the function $\Phi\in\gCPCi$, 
once we have shown that $\Pi$ is definable in the constant-free polynomial hierarchy over $\R$
and that for any fixed instance~$I$ of $\Projchar$, property $\Pi$ is satisfied by almost all $\uF\in \Flag$
(cf. Definition~\ref{def:genpr}).

Lemma~\ref{le:proj-char} tells us that $\varphi\pitchfork \Omega_\lambda(\uF)$ is
satisfied for almost all $\uF\in\Flag$.
In order to show that $\dim R_{\lambda}(\uF) < m-|\lambda|$ for almost all~$\uF$,
we apply Lemma~\ref{le:trans-Schubert} to the quasiprojective set
$W_j:=\{x\in W\mid \dim\psi(x)=j\}$ and the map
$\psi_j\colon W_j\to \Gr(j,n),x\mapsto\psi(x)$, for $0\le j\le m$.
It is not hard to identify the set
$$
 R_{j,\lambda}(\uF) :=\{x\in W_j\mid \dim(\psi(x)\cap F_{\sigma_i}) \ge i\mbox{ for $0\le i \le m$}\}
$$
as the preimage of the Schubert variety corresponding to the flag 
$\uF$ and to a partition $\mu^{(j)}$ satisfying 
$|\mu^{(j)}|\geq |\lambda|$. 
Thus $R_{j,\lambda}(\uF)$ has dimension $\dim W_j-|\mu|\leq \dim W_j -|\lambda|$ for almost all~$\uF$.
Since $W=W_0\cup \dots \cup W_m$ and $\dim W<m$ we have
$R_\lambda(\uF)=R_{0,\lambda}(\uF)\cup\cdots\cup R_{m,\lambda}(\uF)$, and
conclude that indeed $\dim R_{\lambda}(\uF) < m-|\lambda|$.

It remains to be seen that $\Pi$ can be defined in $\PHR^0$.
According to Definition~\ref{eq:def-trans}, $\varphi\pitchfork \Omega_\lambda(\uF)$
can be expressed as follows:  
\begin{equation}\label{eq:transc}
 \forall \mu\ (\mu \supseteq \lambda \land \mu \text{ admissible } \Longrightarrow 
  \varphi \pitchfork e_\mu(\uF) ),
\end{equation}
where the transversality condition $\varphi \pitchfork e_\mu(\uF)$ means that 
\begin{equation*}
  \forall x\ ( x \in V  \wedge \varphi(x)\in e_\mu(\uF)
  \Longrightarrow \varphi\pitchfork_x e_\mu(\uF)).
\end{equation*}

Lemma~\ref{le:trans-ph} in Appendix \ref{se:trans} says that  
the local transversality condition in the parenthesis is decidable in~$\PC^0$. 
This implies that condition~(\ref{eq:transc}) is expressible in $\coNPC^0$ and thus 
in $\PHR^0$. 

In order to express $\dim R_\lambda(\uF)<m-|\lambda|$,
we recall that the points $x\in W$ can be characterized among the points of $V'$ as those having 
local dimension smaller than~$m$, cf. Lemma~\ref{le:input}.
The local dimension of (semi)algebraic sets is expressible in the constant-free polynomial hierarchy
over the reals (compare Lemma~\ref{le:dim-ph}). 
We can thus express membership to $R_\lambda(\uF)$ in $\PHR^0$. 
Finally, using Lemma~\ref{le:dim-ph} again, we conclude that 
the condition $\dim R_\lambda(\uF)<m-|\lambda|$
is expressible in $\PHR^0$.
\endProofof

\subsection{Lower bounds}

We first complement the upper bound in Corollary~\ref{cor:hilbs-ub} by
a lower bound.

\begin{proposition}\label{pro:hilbs-lb}
The problem $\HILBSbit$ is $\CP$-hard.
\end{proposition}

\begin{proof}
We proceed as in~\cite{bach:99}.
Let $\varphi$ be a Boolean formula in the variables $X_1,\ldots,X_n$ in
conjunctive normal form. It is well
known that the problem $\#\mathrm{SAT}$ to count the number of satisfying assignments of
such formulas is $\CP$-complete~\cite{vali:79-1,vali:79-2}.

For each literal $\lambda$ put $g_\lambda:=1-X_i$ if $\lambda=X_i$ and
$g_\lambda:=X_i$ if $\lambda$ is the negation of~$X_i$.
For each clause $\kappa=\lambda_1\vee\cdots\vee\lambda_k$ put
$g_\kappa:=\prod_{i=1}^k g_{\lambda_i}$.
Let $f_\kappa$ denote the homogenization of $g_\kappa$ with respect to
the variable $X_0$.

We assign to the Boolean formula $\varphi=\kappa_1\wedge\cdots\wedge\kappa_s$
the system of homogeneous equations
$$
 X_1^2-X_1X_0,\ldots,X_n^2-X_nX_0,f_{\kappa_1},\ldots,f_{\kappa_s}.
$$
Clearly, the zero set $V'$ of this system in $\PP^n$ corresponds bijectively
to the satisfying assignments of $\varphi$ (there are no solutions at infinity).
Moreover, looking at the first $n$ equations we see that the
input condition~(\ref{eq:input-cond}) is satisfied with $m=0$.
The Hilbert polynomial of $V'$ is constant and equals the number of
satisfying assignments of $\varphi$.
This provides a polynomial time reduction from $\#\mathrm{SAT}$ to $\HILBSbit$.
\end{proof}

\begin{remark}\label{eq:lb-HILBS}
Due to the input condition~(\ref{eq:input-cond})
it is not clear whether $\HILBS$ and $\HILBSbit$ are $\CPCi$-hard
and $\GCC$-hard, respectively.
\end{remark}

Corollary~\ref{cor:hilbs-ub} states that the problem $\HILBSbit$ to
compute the Hilbert polynomial of smooth varieties is in $\BP(\gGapC)$.
We next show that the general problem to compute the Hilbert polynomial
of a homogeneous ideal is presumably more difficult, namely $\FPSPACE$-hard.
Consider the following problems:

\begin{description}
\item{$\HIM$} ({\em Homogeneous ideal membership problem})\quad
Given non-constant homogeneous polynomials
$f_1,\dots,f_r,g\in\C[X_0,\dots,X_n]$, decide whether
$g$ lies in the ideal generated by $f_1,\ldots,f_r$.

\item{$\HILB$} ({\em Hilbert polynomial})\quad Given
a family of non-constant homogeneous polynomials
$f_1,\dots,f_r$ in $\C[X_0,\dots,X_n]$ and $0\le k\le n$,
compute the $k$-th coefficient of the Hilbert polynomial
of the homogeneous ideal generated by $f_1,\ldots,f_r$.
\end{description}

We will use the following simple and well-known lemma to establish a
Turing reduction from $\HIMbit$ to $\HILBbit$, and then invoke
a result in Mayr~\cite[Thm.~17]{mayr:97},
which states that $\HIMbit$ is $\PSPACE$-complete.

\begin{lemma}
Let $I$ be a homogeneous ideal such that some $X_i$ is not a zero-divisor of $\C[X_0,\dots,X_n]/I$.
Let $g$ be a non-constant homogeneous polynomial. Then
$g\in I$ if and only if $I$ and $I+(g)$ have the same Hilbert polynomial.
\end{lemma}

\begin{proof}
Assume $X_i$ is not a zero-divisor of $\C[X_0,\dots,X_n]/I$.
Let $I$, $g$ be such that $J:=I+(g)$ and $I$ have the same Hilbert polynomial.
This means that $J^{(d)}=I^{(d)}$ for sufficiently large degree~$d$.
Hence, we have $X_i^d g\in I$ for sufficiently large $d$, and thus $g\in I$.
\end{proof}

By introducing a further variable $Y$ we can achieve that $Y$ is
not a zero-divisor of $\C[X_0,\dots,X_n,Y]/\overline{I}$, where $\overline{I}=\C[X_0,\dots,X_n,Y]I$.
Hence we obtain the following lower bound.

\begin{theorem}\label{th:hilb-hard}
The problem $\HILBbit$ is $\FPSPACE$-hard.
\end{theorem}

Based on this theorem, we can now improve the $\CP$-lower bound
in~\cite{bach:99} for the problem to compute the ranks of cohomology groups
of coherent sheaves on projective space. The lower bound is also true for the problem to
compute the corresponding Euler characteristic.

For an introduction to sheaf cohomology we refer to~\cite{hart:77,iita:82}.
We encode the input to our problems as in~\cite{bach:99}.
Thus we specify a coherent sheaf on $\PP^n$ by giving a {\em graded matrix}.
This is a matrix $(p_{ij})_{1\le i\le s,1\le j\le r}$
of homogeneous polynomials in $S:=\C[X_0,\ldots,X_n]$ together with
two arrays of integers $(d_1,\ldots,d_s)$ and $(e_1,\ldots,e_r)$ such that
$\deg p_{ij} = d_i -e_j$ whenever $p_{ij}\ne 0$.
A graded matrix defines a degree-preserving morphism
$$
 \gamma \colon\bigoplus_{j=1}^r S(e_j)\to\bigoplus_{i=1}^s S(d_i)
$$
of graded $S$-modules. (As usual, $S(d)$ denotes $S$ with degrees shifted
by $d$ to the left, so that $S(d)_0=S_d$.)
The cokernel $M$ of $\gamma$ is a finitely generated, graded $S$-module and thus determines
a coherent sheaf $\tilde{M}$ on $\PP^n$ (cf.~\cite[p.~116]{hart:77}).
We study the task to compute the dimensions of the
cohomology $\C$-vector spaces $H^i(\PP^n,\tilde{M})$ for $i=0,\ldots,n$.
(It is known that these vector spaces vanish for $i>n$ \cite[III.2.7]{hart:77}.)
The Euler characteristic of the sheaf $\tilde{M}$ is defined as
\begin{equation}\label{eq:sheafeuler}
\chi(\tilde{M}):=\sum_{i=0}^n (-1)^i \dim H^i(\PP^n,\tilde{M}).
\end{equation}
The link to the Hilbert polynomial is given by the following proposition,
a proof of which can be found in \cite[Section 7.6]{iita:82}, see also \cite[Ex. III.5.2]{hart:77}.

\begin{proposition}\label{prop:hileuler}
Let $I\subseteq S:=\C[X_0,\dots,X_n]$ be a homogeneous ideal, $M=S/I$ and $p_M(T)\in \Q[T]$ the corresponding Hilbert polynomial.
Then $p_M(d)=\chi(\tilde{M}(d))$ for all $d\in \Z$.
\end{proposition}

We now consider the following problems.

\begin{description}
\item{$\RankSheaf$} ({\em Rank of sheaf cohomology})\quad
Given a morphism $\gamma$ by a graded matrix as above and given $i\in\N$, compute
$\dim H^i(\PP^n,\tilde{M})$ for $M=\coker\gamma$.

\item{$\EulerSheaf$} ({\em Euler characteristic of sheaf cohomology})\quad
Given a morphism $\gamma$ by a graded matrix as above, compute
$\chi(\tilde{M})$ for $M=\coker\gamma$.
\end{description}

The following result improves the $\CP$-lower bound in~\cite{bach:99}.

\begin{corollary}\label{cor:bettisheaf}
The problems $\RankSheafbit$ and $\EulerSheafbit$ are $\FPSPACE$-hard.
\end{corollary}

\begin{proof}
Clearly, $\EulerSheafbit$ can be Turing reduced to $\RankSheafbit$.
Theorem~\ref{th:hilb-hard} tells us that $\HILBbit$ is $\FPSPACE$-hard.
It is therefore sufficient to establish a Turing reduction from $\HILBbit$
to $\EulerSheafbit$.

An instance of $\HILBbit$ is a family of non-constant homogeneous polynomials
$f_1,\dots,f_r$ in $\Z[X_0,\dots,X_n]$. Let $I$ denote the corresponding
homogeneous ideal in $\C[X_0,\dots,X_n]$. Consider the graded morphism
$\gamma\colon\oplus_{j=1}^r S(e_j)\to S$ given by $f_1,\ldots,f_r$, where
$e_j:=-\deg f_j$. The cokernel $M$ of $\gamma$ equals $S/I$.

By Proposition~\ref{prop:hileuler} we have $p_M(d)=\chi(\tilde{M}(d))$ for all $d\in\Z$.
We can therefore obtain the values $p_M(d)$ for $d=0,\ldots,n$
by $n+1$ calls to $\EulerSheafbit$ and then compute the coefficients of
$p_M$ by interpolation.
\end{proof}

\begin{remark}
The algorithm in~\cite{bast:92} combined with the upper bounds
in \cite{mayr:97} implies that $\HILBbit$ is in $\FEXPSPACE$.
We do not know of any better upper bound on this problem.
The known algorithms for sheaf cohomology (cf.~\cite[Chapter 8]{vasc:98}, \cite{deei:02})
suggest that $\RankSheafbit$ is in $\FEXPSPACE$.
\end{remark}

\section{Hilbert polynomial and degeneracy loci}\label{se:hidegen}

This section is devoted to the proof of Theorem \ref{th:hmain}.

\subsection{Chern classes and Riemann-Roch}\label{se:chern}

References for the material presented
here are \cite{cher:46,hirz:66,mist:74}. See also \cite{fult:98} for
the algebraic geometry perspective.  Let $V$ be a variety (recall the conventions made for varieties
at the beginning of \S\ref{se:prel}).
Chern classes are characteristic cohomology classes
$c_i(E)\in H^{2i}(V)$ associated to a complex vector bundle
$p\colon E\rightarrow V$. Chern classes are characterized axiomatically as follows: 

\begin{enumerate}
\item $c_i(E)\in H^{2i}(V)$, $c_0(E)=1$ and $c_1(\mathscr{L})$ generates $H^2(\PP^n)$, where $\mathscr{L}$ is the
canonical line bundle on $\PP^n$.
\item Let $f\colon W\rightarrow V$ be a morphism of projective varieties. Then $c_i(f^*(E))=f^*(c_i(E))$,
where $f^*(E)$ denotes the pull-back bundle with
respect to $f$.
\item (Whitney formula.) An exact sequence of bundles $0\rightarrow E'\rightarrow E\rightarrow E''\rightarrow 0$,
implies $c_k(E)=\sum_{i}c_i(E')\cupp c_{k-i}(E'')$, where $\cupp$ denotes the cup-product.
\end{enumerate}

The total Chern class is the sum $c(E)=\sum_{i\geq 0}c_i(E)\in H^*(V)$ of all the Chern classes.
If $V$ is smooth and irreducible of dimension $m$, then the top Chern class $c_m(TV)$ of the tangent bundle
evaluated at the fundamental class yields the topological Euler characteristic of $V$:
\begin{equation*}
  \chi(V)=\deg (c_m(TV)\capp [V]),
\end{equation*}
see \cite[page 170]{mist:74}.
Here $\capp$ denotes the cap-product and
$\deg\colon H_0 (V)\to\Z$ is defined by
$\deg(\sum_p n_p\, [p]) = \sum_p n_p$.

We now introduce the necessary terminology needed to state the Hirzebruch-Riemann-Roch theorem, which relates
the Chern classes to the Hilbert polynomial.

Let $f(t)=\frac{t}{1-e^{-t}}\in \Q[[t]]$ be the formal power series (\ref{eq:def-b})
and $t_1,\dots,t_n$ different variables.
Consider the product
\begin{equation*}
  f(t_1)\cdots f(t_n)=\sum_{i=0}^{\infty}g_i(t_1,\dots,t_n),
\end{equation*}
where the $g_i$ are the $i$-th graded parts.
The $g_i$ are symmetric polynomials in the~$t_i$, so there is an expression
$g_n(t_1,\dots,t_n)=T_n(\sigma_1,\dots,\sigma_n)$,
where $\sigma_i$ is the $i$-th elementary symmetric function in the $t_i$.
The $T_n\in \Q[X_1,\dots,X_n]$ are called {\em Todd polynomials}.
Note that $T_n$ is homogeneous of weight $n$, when we define the weight of a monomial 
$X_1^{i_1}\cdots X_n^{i_n}$ to be the sum $\sum_{k=1}^n ki_k$.
For example, the first three Todd polynomials are: $T_1=\frac{1}{2}X_1,\ T_2=\frac{1}{12}(X_1^2+X_2),\ T_3=\frac{1}{24}X_1X_2$.

If $c(TV)$ is the total Chern polynomial of the tangent bundle of a smooth variety~$V$ of dimension $m$, then we
define the Todd class of $V$ to be
\begin{equation*}
  \mathrm{td}(V):=1+\sum_{i=1}^{m}T_i(c_1,\dots,c_i),
\end{equation*}
where here (and later) we write $c_i$ as shorthand for $c_i(TV)$.

Consider the sum $\sum_{i=1}^n e^{t_i}=n+\sum_{i\geq 1} p_i(t_1,\dots,t_n)$.
Again, the $p_i$ are symmetric, so there is a polynomial $K_n(X_1,\dots,X_n)$ which evaluated at the
elementary symmetric functions in the $t_i$ yields $p_n$. If $c_i(E)$ are
the Chern classes of a vector bundle $E$ on a variety $V$, then the class
\begin{equation*}
  \mathrm{ch}(E):=1+\sum_{i\geq 1} K_i(c_1(E),\dots,c_i(E))
\end{equation*}
is called the {\em Chern character} of $E$.

To a variety $V\subseteq\PP^n$ and $d\in\Z$ one can assign the twisted sheaf $\Oh_V(d)$.
The Chern character of the sheaf $\Oh_V(d)$ is particularly easy to describe.
Since $\Oh_V(d)$ corresponds to a line bundle, we have only a first Chern class,
which is $c_1(\Oh_V(d))=d c_1(\mathscr{L}_V)$.
Here, and in what follows, $\mathscr{L}_V$ denotes the line bundle corresponding to
the sheaf $\Oh_V(1)$ and $\mathscr{L}_V^{\vee}$ its dual, i.e., the canonical line bundle on $V$.
For the Chern character we get
\begin{equation}\label{eq:cchar-lb}
\mathrm{ch}(\Oh_V(d))=e^{c_1(\Oh_V(d))}=\sum_{i\geq 0} \frac{d^i}{i!}c_1(\mathscr{L}_V)^i.
\end{equation}

To a vector bundle $E$ on a variety $V$ there corresponds a locally free sheaf $\mathscr{E}$,
see \cite[VI.1.3]{shaf:74} or \cite[Ex. II.5.18]{hart:77}.
Thus we can define the Euler characteristic $\chi(E)$ of~$E$ to be the Euler characteristic
$\chi(\mathscr{E})=\sum_i (-1)^i \dim H^i(V,\mathscr{E})$ of $\mathscr{E}$
with respect to sheaf cohomology, cf. Equation~(\ref{eq:sheafeuler}).

\begin{lemma}\label{le:euler-hilb}
Let  $V\subseteq \PP^n$ be a variety and $d\in\Z$. Then
the Euler characteristic of the line bundle $\Oh_V(d)$ equals
the Hilbert polynomial of $V$ evaluated at $d$, that is, $\chi(\Oh_V(d))=p_V(d)$.
\end{lemma}

\begin{proof}
Let $i\colon V\rightarrow \PP^n$ be the inclusion and $\mathscr{F}$ be a coherent sheaf on $V$.
Then $H^i(V,\mathscr{F})=H^i(\PP^n,i_*\mathscr{F})$ for all $i$, cf.~\cite[Lemma III.2.10]{hart:77}.
If $M=S/I$ denotes the homogeneous coordinate ring of $V$, then $i_*\Oh_V(d)=\tilde{M}(d)$,
so by Proposition \ref{prop:hileuler} we have $\chi(\Oh_V(d))=p_V(d)$.
\end{proof}

With all these notions introduced, we can formulate the Hirzebruch-Riemann-Roch theorem.

\begin{theorem}[Hirzebruch-Riemann-Roch, \cite{hirz:66}]\label{thm:hirzebruch}
Let $E$ be a vector bundle on an irreducible smooth variety $V$ of dimension $m$.
Then
\begin{equation*}
  \chi(E)=\deg \left((\mathrm{ch}(E)\cupp \mathrm{td}(V))_m \capp [V]\right).
\end{equation*}
\end{theorem}

Theorem~\ref{thm:hirzebruch} combined with Lemma~\ref{le:euler-hilb} and Equation (\ref{eq:cchar-lb}) immediately yields the
following.

\begin{corollary}\label{co:hilhrr}
Let $V\subseteq \PP^n$ be an irreducible, smooth variety of dimension~$m$.
Then the $k$-th coefficient of the Hilbert polynomial of $V$ is given by
\begin{equation*}
\label{eq:hrr}
  p_k(V)=\frac{1}{k!}\deg (c_1(\mathscr{L}_V)^k\cupp T_{m-k}\left(c_1,\dots,c_{m-k} \right)\capp [V]),
\end{equation*}
where $c_1,\dots,c_m$ are the Chern classes of the tangent bundle $TV$.
\end{corollary}

\subsection{Generalities on symmetric functions}
We gather some results from the theory of symmetric functions that will be used later.
Reference for this material are \cite{macd:95,mani:01}, \cite[Appendix A]{fuha:91} and \cite[Appendix A.9]{fult:98}.

For a partition $\lambda$, we denote by $\lambda'$ its conjugate partition.
Recall the definition of $\Delta_{\lambda}(c)$ in Equation (\ref{eq:defDelta}).
Claims 1 and 2 of the following lemma are easy to verify, a proof of the third one is given in \cite[Lemma A.9.2]{fult:98}.

\begin{lemma}\label{le:deltalambda}
Let $\lambda$ be a partition and $c=\{c_i\}_{i\in \N}$ be a sequence of elements
of a commutative ring such that $c_0=1$.
\begin{enumerate}
\item The polynomial $\Delta_\lambda(c)$ is homogeneous of weight $|\lambda|$ in the $c_i$, when $c_i$ has weight~$i$.
\item Let $c^{\vee}=\{(-1)^ic_i\}_{i\in \N}$. Then $\Delta_{\lambda}(c^{\vee})=(-1)^{|\lambda|}\Delta_{\lambda}(c)$.
\item Let $c^{-1}=\{c_i'\}_{i\in \N}$,  where the $c_i'$ are the coefficients of the
inverse power series $(\sum_{i\geq 0}c_it^i)^{-1}$. Then $\Delta_{\lambda}(c^{-1})=\Delta_{\lambda'}(c^{\vee})$.
\end{enumerate}
\end{lemma}

\begin{example}
We verify claims 2 and 3 of the previous lemma for the special case $\lambda=(1^k)$.
For the partition $(k)$ we have $\Delta_{(k)}(c)=c_k$.
For the partition $(1^k)$ we have $\Delta_{(1^k)}(c)=\det M_k(c)$, where $M_k(c)$ is the Toeplitz matrix
\begin{equation*}
  M_k(c)= \begin{pmatrix}
                         c_{1} & c_{2} & c_3 & \cdots & c_{k-1} & c_{k}\\
                         1 & c_{1} & c_2 & \cdots & c_{k-2} & c_{k-1}\\
                         0 & 1 & c_1 & \cdots & c_{k-3} & c_{k-2}\\
                         \cdots & \cdots & \cdots & \cdots & \cdots & \cdots \\
                         0 & 0 & 0 & \cdots & 1 & c_{1}
                       \end{pmatrix},
\end{equation*}
We can expand the determinant as
\begin{equation*}
  \det M_k(c)=-\sum_{i=1}^k (-1)^i c_i \det M_{k-i}(c).
\end{equation*}
This equation coincides with the recursive formula for the coefficients $c_j'$
of the inverse power series $(\sum_{i\geq 0}(-1)^ic_i)^{-1}$. In particular, we obtain
$\Delta_{(1^k)}(c)=\det M_k(c)=c_k'=\Delta_{(k)}\left(c^{-1}\right)$.
\end{example}

Let $\gamma=(\gamma_1,\dots,\gamma_m)$ be variables and $\lambda$ be a partition such that $|\lambda|\leq m$. 
Define the Schur polynomial associated to $\lambda$ as
\begin{equation}\label{eq:schurdef}
  s_\lambda(\gamma):=s_\lambda(\gamma_1,\dots,\gamma_m)=
\frac{\det (\gamma_i^{\lambda_j+m-j})_{1\leq i,j\leq m}}{\det (\gamma_i^{m-j})_{1\leq i,j\leq m}}.
\end{equation}
The polynomial $s_\lambda(\gamma)$ is symmetric and homogeneous of degree $|\lambda|$. 
Note that $s_\lambda$ depends not only on the partition~$\lambda$ but also on~$m$. 

A proof of the following lemma can be found in \cite[I.3]{macd:95} and \cite[Appendix A]{fuha:91}.

\begin{lemma}[Giambelli's formula]\label{lm:giambelli}
Let $\lambda$ be a partition with $|\lambda|\leq m$ and $c=\{c_i\}_{i\in \N}$ be given such that
\begin{equation*}
  c_0+c_1t+\dots +c_mt^m=\prod_{i=1}^m (1+\gamma_it),
\end{equation*}
i.e., the $c_i$ are elementary symmetric functions in the $\gamma_j$.
Then $\Delta_\lambda(c)=s_{\lambda'}(\gamma)$.
\end{lemma}

\begin{example}\label{ex:Schur-spezial}
If $\lambda=(k)$, then $s_\lambda(\gamma)$ is the $k$-th complete symmetric polynomial in the $\gamma_i$.
This is the sum of all distinct monomials of degree $k$ in the $\gamma_i$.
If $\lambda=(1^k)$, then $s_\lambda(\gamma)$ is the $k$-th elementary symmetric function in the $\gamma_i$.
\end{example}

We will further need a formula expanding the Schur polynomial of a sum of variables. The following lemma
follows from \cite[Example A.9.1]{fult:98} (see also \cite[Example I.3.10]{macd:95}).

\begin{lemma}\label{le:expand-Schur}
Let $\lambda$ be a partition with $|\lambda|\leq m$. Then
\begin{equation*}
 s_{\lambda}(\gamma_1+\beta,\dots,\gamma_{m+1}+\beta)=\sum_{\mu\subseteq \lambda}
d_{\lambda \mu}^m\beta^{|\lambda|-|\mu|} s_{\mu}(\gamma_1,\dots,\gamma_{m+1}),
\end{equation*}
where
\begin{equation*}
d_{\lambda \mu}^m=\det \binom{\lambda_i+m+1-i}{\mu_j+m+1-j}_{1\leq i,j\leq m}
\end{equation*}
\end{lemma}

\begin{example}\label{ex:gaga}
Let $\lambda=(1^k)$. Then any subpartition $\mu\subseteq \lambda$ is of the form $(1^j)$ for some
$j\leq k$ and $d_{\lambda \mu}^m=\binom{m-j+1}{m-k+1}$. This follows from looking at the coefficients
of the expansion of $s_{(1^k)}(\gamma_1+\beta,\dots,\gamma_{m+1}+\beta)$, using the fact that $s_{(1^k)}$
is an elementary symmetric function (see Example \ref{ex:Schur-spezial}).
\end{example}

\subsection{Proof of Theorem \ref{th:hmain}}
In this section we derive Theorem \ref{th:hmain} from Corollary \ref{co:hilhrr} in a series of reductions.
We start by observing a determinantal formula for the Todd polynomials. For what follows, we will often
write $T_m(c)$ as shorthand for $T_m(c_1,\dots,c_m)$.

\begin{lemma}
\label{lm:schur}
Let $b=\{b_i\}_{i\in \N}$ be the sequence of rational numbers from Equation~(\ref{eq:def-b}), 
$c_0=1$ and let $c_1,\dots,c_m$ be variables.
Then the $m$-th Todd polynomial is given by
\begin{equation*}
\label{eq:schur}
  T_m(c)=\sum_{|\lambda|=m}\Delta_{\lambda'}(b)\Delta_{\lambda}(c),
\end{equation*}
where $\lambda=(\lambda_1,\dots,\lambda_m)$ runs over all partitions of $m$.
\end{lemma}

\begin{Proof}
Consider the (formal) factorizations
\begin{equation*}
  1+\sum_{i=1}^m c_i=\prod_{j=1}^m (1+\gamma_j), \ 1+\sum_{i=1}^m b_i=\prod_{j=1}^m (1+\beta_j).
\end{equation*}
This amounts to writing the $c_i$ and $b_i$ as elementary symmetric functions in the $\gamma_j$ and $\beta_j$,
respectively.
For a partition $\lambda$ of $m$, let $m_{\lambda}(\gamma)$ be the sum of all different monomials arising from
$\gamma_1^{\lambda_1}\cdots \gamma_m^{\lambda_m}$ by permutation of the $\gamma_i$ (for example,
$m_{(1^m)}(\gamma)=\gamma_1\cdots \gamma_m$). Also, let
$\sigma_{\lambda}=\sigma_{\lambda_1}\cdots \sigma_{\lambda_m}$ denote the product of the elementary symmetric functions
indexed by the partition.
By definition,
$T_m(c)$ is the $m$-th graded component of $f(\gamma_1)\cdots f(\gamma_m)$, where
$f(\gamma_i)=\sum_{j\geq 0}b_j\gamma_i^j$.
It follows that
\begin{align*}
  T_m(c)&=\sum_{i_1+\dots+i_m=m}b_{i_1}\cdots b_{i_m}\gamma_1^{i_1}\cdots \gamma_m^{i_m}\\
  &=\sum_{|\lambda|=m}b_{\lambda_1} \cdots b_{\lambda_m}m_{\lambda}(\gamma)
=\sum_{|\lambda|=m}\sigma_{\lambda}(\beta)m_{\lambda}(\gamma).
\end{align*}
By \cite[I.4(4.2'-3')]{macd:95} we have
\begin{equation}\label{eq:macd}
  \sum_{|\lambda|\leq m}\sigma_{\lambda}(\beta)m_{\lambda}(\gamma)=\prod_{1\leq i,j\leq m}(1+\beta_j\gamma_i)=
 \sum_{|\lambda|\leq m}s_{\lambda'}(\beta)s_{\lambda}(\gamma),
\end{equation}
where $s_{\lambda}$ is the Schur polynomial of the partition $\lambda$. Giambelli's formula (Lemma~\ref{lm:giambelli})
expresses the Schur polynomials as determinants:
\begin{equation*}
  s_{\lambda}(\gamma)=\Delta_{\lambda'}(c).
\end{equation*}
Noting that $\deg s_{\lambda}(\gamma)=\deg m_{\lambda}(\gamma)=|\lambda|$ and taking the degree $m$
parts in (\ref{eq:macd}) completes the proof.
\end{Proof}

What makes this formula useful is the fact that if $c$
denotes the total Chern class of the tangent bundle of
a smooth variety, the cohomology classes $\Delta_{\lambda}(c)$ can be put in relation
to homology classes $[P_\lambda]$ of the generalized polar varieties.

To a smooth variety $V\subseteq \PP^n$ of dimension $m$ we can associate a vector bundle $\tilde{T}V$ of rank $m+1$ such
that for all $x\in V$, $\mathbb{T}_xV=\PP(\tilde{T}_xV)$.

The proof of the following proposition uses a result of Kempf and Laksov \cite[Theorem 10]{kela:74},
see also \cite[Theorem 14.3]{fult:98}.

\begin{proposition}
\label{prop:fult14}
Let $V\subseteq \PP^n$ be an irreducible, smooth variety of dimension $m$ and let $\lambda$ be a partition
with $|\lambda|\leq m$. Then
\begin{equation*}
  \Delta_{\lambda'}(c(\tilde{T}V))\capp [V]=\begin{cases}
    (-1)^{|\lambda|} [P_{\lambda}] & \text{ if } \lambda_1\leq n-m\\
    0 & \text{ else}.
  \end{cases}
\end{equation*}
\end{proposition}

\begin{Proof}
Let $E$ be the trivial $(n+1)$-bundle on $V$, $\tilde{N}V:=E/\tilde{T}V$ and
$\pi\colon E\rightarrow \tilde{N}V$ be the projection map. Let $\lambda$ be a partition with
$|\lambda|\leq m$ and $\lambda_1\leq n-m$.
A flag $\uF\in \Flag$ determines a partial
flag $\underline{A}$ of trivial subbundles of $E$ with $A_i$ corresponding to $F_{\sigma_{i-1}}$ for
$1\leq i\leq m$. Thus $\mathrm{rank}(A_{i})=\sigma_{i-1}+1=n-m+i-\lambda_i$.
The determinantal locus
\begin{equation*}
  \Omega_{\lambda}(\underline{A};\pi)=\{ x\in V \ | \ \dim (\ker \pi (x)\cap A_i(x))\geq i, \ 1\leq i\leq m\}
\end{equation*}
studied in \cite[Chapter 14]{fult:98} coincides with the generalized polar variety $P_{\lambda}(\uF)$.
Here, by $\dim (\ker \pi (x)\cap A_i(x))$ we mean the affine dimension.
The statement of \cite[Theorem 14.3]{fult:98} implies that
\begin{equation*}
  \Delta_\lambda(c(\tilde{N}V))\capp [V]=[\Omega_\lambda(\underline{A};\pi)]=[P_{\lambda}(\uF)],
\end{equation*}
provided $\Omega_\lambda(\underline{A};\pi)$ is of pure codimension $|\lambda|$.
For justifying this, note that
in \cite{fult:98}, $\Omega_\lambda(\underline{A};\pi)$ is interpreted as a subscheme of $V$ and
its class is an element of the Chow group
$A_*(\Omega_\lambda(\underline{A};\pi))$. However, for generic $\uF\in \Flag$,
the scheme $\Omega_\lambda(\underline{A};\pi)$ is
multiplicity-free and of the right codimension, cf.~Lemma \ref{le:proj-char}.
Moreover, there is a cycle map
$A_*(\Omega_\lambda(\underline{A};\pi))\rightarrow H_*(\Omega_\lambda(\underline{A};\sigma))$
\cite[Chapter 19]{fult:98},
which is compatible with the action of Chern classes.

Let $s(\tilde{T}V):=1/c(\tilde{T}V)$ and $\tilde{T}V^{\vee}$ denote the dual bundle.
We have $c(\tilde{N}V)=s(\tilde{T}V)$ and $c_i(\tilde{T}V^{\vee})=(-1)^i c_i(\tilde{T}V)$ \cite{fult:98}.
Using Lemma \ref{le:deltalambda}, we thus get
\begin{equation*}
   \Delta_\lambda(c(\tilde{N}V))=\Delta_{\lambda}(s(\tilde{T}V))
=\Delta_{\lambda'}(c(\tilde{T}V^{\vee}))=(-1)^{|\lambda|}\Delta_{\lambda'}(c(\tilde{T}V)).
\end{equation*}
This shows the assertion in the case $\lambda_1\leq n-m$.
If $\lambda_1>n-m$, then since $\tilde{N}V$ is an $(n-m)$-bundle, we have $c_{j}(\tilde{N}V)=0$ for $j\geq \lambda_1$,
which in turn implies $\Delta_\lambda(c(\tilde{N}V))=0$. This completes the proof.
\end{Proof}

We now turn attention to the tangent bundle $TV$.

\begin{lemma}
\label{le:tensor}
Let $V\subseteq \PP^n$ be an irreducible, smooth variety of dimension~$m$ and let
$\lambda$ be a partition with $|\lambda|\leq m$. For the tangent bundle $TV$ we have
\begin{equation*}
\label{eq:tangent}
  \Delta_{\lambda'}(c(TV))\capp [V]=\sum_{\substack{ \mu\subseteq \lambda\\ \mu_1\leq n-m}}
(-1)^{|\mu|}d_{\lambda \mu}^{m}c_1(\mathscr{L}_V)^{|\lambda|-|\mu|}\capp [P_\mu],
\end{equation*}
where $d_{\lambda \mu}^m$ is defined as in Lemma \ref{le:expand-Schur}.
\end{lemma}

\begin{Proof}
It is well known (compare \cite[Chapter 16]{harr:92}) that the tangent bundle $TV$ of a smooth variety is given by
$TV\cong \mathrm{Hom}\left( \mathscr{L}_V^{\vee},\tilde{T}V/\mathscr{L}_V^{\vee}\right).$ 
Taking the direct sum with the trivial bundle $E=\mathrm{Hom}(\mathscr{L}_V^{\vee},\mathscr{L}_V^{\vee})$ we get
\begin{equation*}
  TV\oplus E\cong \mathrm{Hom}\left( \mathscr{L}_V^{\vee},\tilde{T}V/\mathscr{L}_V^{\vee}\right)\oplus E
  \cong \mathrm{Hom}(\mathscr{L}_V^{\vee},\tilde{T}V)\cong \mathscr{L}_V\otimes \tilde{T}V.
\end{equation*}
By the Whitney product formula (see \S\ref{se:chern}) and the fact that $c(E)=1$ we obtain
\begin{equation*}
  c(TV)=c(TV\oplus E)=c(\mathscr{L}_V\otimes \tilde{T}V).
\end{equation*}
Let $c(\tilde{T}V)=\prod_{i=1}^{m+1}(1+\gamma_i)$ be the formal factorization and set
$\beta:=c_1(\mathscr{L}_V)$. 
By Giambelli's formula (Lemma \ref{lm:giambelli})
we have for a partition $\mu$ with $|\mu|\le m$ 
\begin{equation}\label{eq:deltaS}
 \Delta_{\mu'}(c(\tilde{T}V))=s_{\mu}(\gamma_1,\dots,\gamma_{m+1}). 
\end{equation}
On the other hand, it is known that (see \cite[Remark 3.2.3b]{fult:98})
\begin{equation*}
  c(\mathscr{L}_V\otimes \tilde{T}V)=\prod_{i=1}^{m+1}(1+\gamma_i+\beta).
\end{equation*}
Using Lemma \ref{lm:giambelli} again, we get for any partition $\lambda$ with $|\lambda|\leq m$
\begin{equation*}
  \Delta_{\lambda'}(c(TV))  
  = \Delta_{\lambda'}(c(\mathscr{L}_V\otimes \tilde{T}V))
  = s_{\lambda}(\gamma_1+\beta,\dots,\gamma_{m+1}+\beta).
\end{equation*}
By Lemma \ref{le:expand-Schur} we have
\begin{equation*}
  s_{\lambda}(\gamma_1+\beta,\dots,\gamma_{m+1}+\beta)=\sum_{\mu\subseteq \lambda}
d_{\lambda \mu}^m\beta^{|\lambda|-|\mu|}s_{\mu}(\gamma_1,\dots,\gamma_{m+1}).
\end{equation*}
Proposition \ref{prop:fult14} and (\ref{eq:deltaS})
now imply for a partition $\mu$ with $|\mu|\leq m$:
\begin{equation*}
  s_{\mu}(\gamma)\capp [V]=\Delta_{\mu'}(c(\tilde{T}V))\capp [V]=\begin{cases} (-1)^{|\mu|}[P_{\mu}] & \text{ for } \mu_1\leq n-m\\
    0 & \text{ else. }
    \end{cases}
\end{equation*}
This finishes the proof.
\end{Proof}

\begin{example}
Let $\lambda=(1^k)$. Then $\Delta_{\lambda'} (c)=c_k(TV)$, the $k$-th Chern class of the tangent bundle.
By Example \ref{ex:gaga}  we have $d_{\lambda \mu}^m=\binom{m-j+1}{m-k+1}$, 
where $\mu=(1^{j})$.
Plugging this into the formula of Lemma \ref{le:tensor}, we get
\begin{equation*}
c_k(TV)\capp [V]=\sum_{j=0}^k (-1)^j \binom{m-j+1}{m-k+1} c_1(\mathscr{L}_V)^{k-j}\capp [P_j],
\end{equation*}
where the $[P_j]$ are the homology classes of the polar varieties.
This formula is just the known expression for Chern classes in terms of polar classes, see for example \cite{pien:78,bras:00}.
\end{example}

\Proofof{Theorem \ref{th:hmain}}
Assume first that $V$ is irreducible and write $c=c(TV)$.
We express the Poincar\'e dual of the Todd polynomials in terms of the degeneracy loci,
using Lemma~\ref{lm:schur} and Lemma~\ref{le:tensor}:
\begin{align*}
  T_{m-k}(c)\capp [V] &= \sum_{|\lambda|=m-k}\Delta_{\lambda}(b)\Delta_{\lambda'}(c)\capp [V]\\
                      &= \sum_{|\lambda|=m-k}\Delta_{\lambda}(b)\sum_{\substack{\mu\subseteq \lambda \\ \mu_1\leq n-m}}(-1)^{|\mu|}
                      d_{\lambda\mu}^m c_1(\mathscr{L}_V)^{m-k-|\mu|}\capp [P_\mu]\\
                      &= \sum_{\substack{|\mu|\leq m-k\\ \mu_1\leq n-m}}\underbrace{(-1)^{|\mu|}\left(\sum_{\substack{\mu\subseteq \lambda\\
                            |\lambda|=m-k}}
                        \Delta_{\lambda}(b)d_{\lambda \mu}^m\right)}_{=:\delta_{\mu}^{m,k}}
c_1(\mathscr{L}_V)^{m-k-|\mu|}\capp [P_\mu]
\end{align*}
(Recall the definition of $\delta_{\mu}^{m,k}$ in Equation (\ref{eq:defdelta}).)
By Corollary \ref{co:hilhrr} we obtain for the $k$-th coefficient $p_k(V)$ of the Hilbert polynomial of $V$
\begin{align*}
  p_k(V)&=\frac{1}{k!}\deg \left( c_1(\mathscr{L}_V)^k\cupp T_{m-k}\left(c \right)\capp [V]\right)\\
        &=\sum_{\substack{|\mu|\leq m-k\\ \mu_1\leq n-m}}\delta_{\mu}^{m,k} \deg \left(c_1(\mathscr{L}_V)^{m-|\mu|}\capp [P_\mu]\right).
\end{align*}
Since capping with $c_1(\mathscr{L}_V)^{m-|\mu|}$ corresponds to an intersection with a generic linear subspace of
codimension $|\mu|$ in $V$,
we have $\deg \left(c_1(\mathscr{L}_V)^{m-|\mu|}\capp [P_\mu]\right)=\deg P_\mu$. This proves the claim for irreducible $V$.

Now let $V=V_1\cup \dots \cup V_s$ be the decomposition of $V$ into irreducible components of the same dimension.
Let $P_\mu^i$ denote the degeneracy locus of $V_i$ corresponding to~$\mu$ and a generic flag $\uF$.
Since $V$ is smooth, the $V_i$ are pairwise disjoint and
$P_\mu=P_\mu^1\cup \dots \cup P_\mu^s$, 
from which $\deg P_\mu=\sum_i \deg P_\mu^i$ follows.
On the other hand, the Hilbert polynomial is additive on the $V_i$,
which finishes the proof.
\endProofof



\section*{Appendix}\label{se:appendix}

\setcounter{section}{1}
\setcounter{subsection}{0}
\setcounter{theorem}{0}
\renewcommand{\thesection}{\Alph{section}}

\subsection{Proofs of Lemmas \ref{le:proj-char} and \ref{le:trans-Schubert}}
\label{se:app-A1}

For Lemma \ref{le:proj-char} we need the following result of Kleiman~\cite{klei:74},
see also \cite[III.10]{hart:77}.

\begin{lemma}\label{le:kleiman}
Let $\varphi\colon V\rightarrow Y$ be a morphism of smooth irreducible varieties and let
$X\subseteq Y$ be a quasiprojective smooth subvariety. Assume that~$Y$ is a homogeneous space, 
with a connected algebraic group $G$ acting transitively on it.
Then for almost all $g\in G$, $\varphi$ meets $gX$ transversely.
Moreover, if $\delta:=\dim\varphi(V)+\dim X -\dim Y\ge 0$, then 
$\varphi(V)\cap gX$ is of pure dimension~$\delta$, 
for almost all $g\in G$.
\end{lemma}

Recall that a partition $\lambda$ was named admissible if $\lambda_1\leq n-m$ and $|\lambda|\leq m+1$.

\begin{corollary}\label{le:schub-inter}
Let $Z\subseteq\Gr(m,n)$ be a quasiprojective irreducible subvariety
and~$\lambda$ be an admissible partition.
Then, for almost all $\uF\in\Flag$, the intersection $Z\cap\Omega_\lambda(\uF)$ has codimension
$|\lambda|$ in $Z$ if $|\lambda|\le\dim Z$, and it is empty otherwise.
\end{corollary}

\begin{proof}
Recall from~(\ref{eq:Sdecomp}) the cell decomposition 
$\Omega_\lambda(\uF) = \cup_{\lambda\subseteq\mu} e_\mu(\uF)$. 
The Grassmannian $\Gr(m,n)$ is a homogeneous space with respect to the natural action of
the linear group $G:=\GL(n+1,\C)$. 
The group $G$ also acts transitively on
the flag variety~$\Flag$ (in fact, we can define $\Flag$ as a quotient of~$G$,
cf.~\cite[\S 3.6]{mani:01}) and we have $g e_\lambda(\uF)= e_\lambda(g\uF)$.
Decompose $Z$ as finite disjoint union of smooth irreducible quasiprojective varieties $Z_j$. 
We can then apply Lemma~\ref{le:kleiman} to the inclusion of $Z_j$ in $\Gr(m,n)$ and 
to a Schubert cell $X:=e_\mu(\uF)$ in order to obtain that, for almost all $\uF$, 
the  intersection $Z_j\cap e_\mu(\uF)$ has the expected dimension 
(namely $\dim Z_j - |\mu|$ if this is nonnegative, otherwise the intersection is empty).   
 This implies the assertion.
\end{proof}

\Proofof{Lemma \ref{le:proj-char}} 
Without lack of generality we may assume that $V$ is irreducible and not linear.
(Note that for linear $V$, $\varphi(V)$ consists of one point only and thus 
the transversality condition $\varphi \pitchfork \Omega_\lambda(\uF)$
is equivalent to $\varphi(V)\cap \Omega_\lambda(\uF)=\emptyset$, 
except for the trivial case $\lambda=(0)$. 
We may thus safely ignore linear components and restrict attention to a single nonlinear component.)

In this case, a result of Zak~\cite[\S 7]{fula:81} says that
the Gauss map $\varphi\colon V\to\Gr(m,n)$ is finite, hence $\dim \varphi(V)=\dim V=m$. 
Since we are dealing with projective varieties, we have 
$\dim (\varphi(V) \cap \Omega_\lambda(\uF))\ge \dim\varphi(V)+\dim \Omega_\lambda(\uF) -\dim \Gr(m,n) = m-|\lambda|$  
for any partition $\lambda$ with $|\lambda|\le m$  
by a standard dimension argument, cf.~\cite[Thm.17.24]{harr:92}. 

(i)  
Let $\mu\supseteq \lambda$ be an admissible partition. 
Lemma~\ref{le:kleiman} implies that for almost all flags $\uF\in \Flag$, 
$\varphi$ meets $e_\mu(\uF)$ transversely. 
Looking at the cell decomposition~(\ref{eq:Sdecomp}) of $\Omega_\lambda(\uF)$, 
the claim follows (recall Definition~\ref{eq:def-trans}). 
 
(ii) 
We proceed by induction on the size of $\lambda$. Assume that the claim 
is true for all partitions $\mu$ such that $|\lambda|<|\mu|\le m$. 
Suppose $\varphi\pitchfork \Omega_\lambda(\uF)$. 
The cell decomposition~(\ref{eq:Sdecomp}) of $\Omega_\lambda(\uF)$ implies that 
$$
\varphi(V) \cap \Omega_\lambda(\uF) = \bigcup_{\mu\supseteq\lambda} \varphi(V)\cap e_\mu(\uF) . 
$$ 
We are going to show that $\varphi(V)$ intersects the cell $e_\lambda(\uF)$. 
If this were not the case, we had 
$\dim (\varphi(V) \cap \Omega_\lambda(\uF)) = \max_{\mu\supset\lambda} (m -|\mu|) < m -|\lambda|$, 
since we have $\dim (\varphi(V) \cap e_\mu(\uF)) = m -|\mu|$ by induction hypothesis. 
However, this contradicts the fact that 
$\dim (\varphi(V) \cap \Omega_\lambda(\uF))\ge m-|\lambda|$.  

Now note that 
$P_\lambda(\uF) = \cup_{\mu\supseteq\lambda}\varphi^{-1}(e_\lambda(\uF))$.  
By Remark \ref{re:trans}, $\varphi^{-1}(e_\mu(\uF))$ is either empty or of codimension $m-|\mu|$ in $V$. 
Moreover, we just showed that $\varphi^{-1}(e_\lambda(\uF))$ is nonempty.
This show the induction claim. 
The induction start where $|\lambda|=m$ is proved similarly.

(iii) We fix a flag $\uF_0\in\Flag$ and
set $\Omega:=\Omega_\lambda(\uF_0)$, $e:=e_\lambda(\uF_0)$, 
$\partial e:= \Omega - e$. 
Consider the map
$$
 \delta\colon G\rightarrow \N,\ g\mapsto \deg \varphi^{-1}(g\Omega).
$$
It is easy to see that the fibers of $\delta$ are constructible.
Since~$G$ is irreducible, there exists a unique integer $d_\lambda$ such that
$\delta(g)=d_\lambda$ for almost all $g\in G$.
We have to show that
$$
 \forall g\in G \ (\varphi\pitchfork g\Omega \Longrightarrow \delta(g)=d_\lambda ).
$$
Fix $g'\in G$ such that $\varphi\pitchfork g'\Omega$ holds and
write $N:=\delta(g')$. 
By~(ii) we know that $\varphi^{-1}(g'\Omega)$ is of codimension~$|\lambda|$ in~$V$.
It is sufficient to show that the function $\delta$ is constant in a
{\em Euclidean} neighborhood of~$g'$.

Let $A\subseteq \PP^n$ be a linear subspace of dimension $k:=n-m+|\lambda|$ such that
\begin{equation}\label{eq:aphi}
 A \cap \varphi^{-1}(g'\partial e) =\emptyset\ \mbox{ and }
 A \pitchfork \varphi^{-1}(g'e).
 \end{equation}
Then the intersection $A\cap \varphi^{-1}(g'\Omega)$ consists of exactly~$N$ elements,
say $x_1,\ldots,x_N$, cf.~\cite[\S5A]{mumf:76}.
It is therefore sufficient to show that for all $g$ in some neighborhood of~$g'$
condition~(\ref{eq:aphi}) holds with $g$ instead of $g'$ and
$|A\cap\varphi^{-1}(g\Omega)| = N$.

Fix a point $x_i$.
Since $\varphi^{-1}(g'e)$ is smooth and of codimension~$k$ in $\PP^n$,
it can be defined locally around $x_i$ by $k$ equations
$h_1(x,g'),\dots,h_k(x,g')$. Moreover, these equations can be chosen such that
$h_1,\dots,h_{n-m}$ are local equations for $V$ around~$x_i$ (not depending on $g'$) and
$h_{n-m+1},\dots,h_k$ are obtained by pulling back local equations for $g'e$
at the smooth point $\varphi(x)$.
Note that these last $|\lambda|$ equations are polynomials in $x$
as well as in the parameter $g'$.
Suppose that $A$ is the zero set of linear forms $a_1,\dots,a_{n-k}$.
The transversality condition
$A\pitchfork_{x_i} \varphi^{-1}(g'e)$ implies that
$d_xh_1(x_i,g'),\dots,d_xh_k(x_i,g'),a_1,\dots,a_{n-k}$
are linearly independent. We are thus in the situation of the implicit function theorem:
there is a Euclidean neighborhood~$U$ of $g'$ and
a Euclidean neighborhood~$V_i$ of $x_i$ such that for each $g\in U$
the set $A\cap \varphi^{-1}(g\Omega)\cap V_i$ consists of exactly one point~$x_i(g)$.

It remains to be seen that for $g$ sufficiently close to $g'$, the set
$A\cap \varphi^{-1}(g\Omega)$ cannot have more than~$N$ elements.
Suppose by contradiction that there is a sequence
$g_\nu$ in $G$ converging to $g'$ such that for all~$\nu$,
$A\cap \varphi^{-1}(g_\nu \Omega)$ contains a point $y_\nu$ different from
$x_1(g_\nu),\ldots,x_N(g_\nu)$. Since $V$ is compact, by passing to a subsequence,
we may assume that $y_\nu$ converges to a point~$y\in V$.
By continuity, $y \in A\cap \varphi^{-1}(g'\Omega)$, hence
$y=x_i$ for some~$i$. We conclude that $y_\nu = x_i(g_\nu)$ for $\nu$
sufficiently large, contradicting our assumption.
\endProofof

\Proofof{Lemma \ref{le:trans-Schubert}}
We may assume without loss of generality that $W$ is irreducible and
that $\dim\psi^{-1} (\psi(x))$ is constant for $x\in W$, say equal to~$\delta$.
(Decompose $W$ into the locally closed subsets
$W_i:=\{x\in W\mid \dim\psi^{-1} (\psi(x)) =i\}$
and apply the assertion to the irreducible components of $W_i$.)
By \cite[\S I.6.3 Thm.~7]{shaf:74} (see also \cite[Thm.~11.12]{harr:92}) we have
$$
 \dim W = \dim Z +\delta,\quad
 \dim R_\lambda(\uF) \leq \dim\psi(R_\lambda(\uF)) + \delta,
$$
where we have set $Z:=\psi(W)$.
Assume first that $|\lambda|\le\dim Z$.
By Corollary \ref{le:schub-inter}, we have
$\dim(Z\cap \Omega_\lambda(\uF)) = \dim Z - |\lambda|$
for almost all $\uF\in\Flag$. Since
$\psi(R_\lambda(\uF)) = Z \cap \Omega_\lambda(\uF)$,
we obtain for almost all $\uF$
$$
 \dim R_\lambda(\uF) \leq \dim\psi(R_\lambda(\uF)) + \delta =  \dim Z - |\lambda| + \delta
  = \dim W - |\lambda| .
$$
If $|\lambda|>\dim Z$ we have $Z\cap \Omega_\lambda(\uF)=\emptyset$
and therefore $R_\lambda(\uF)=\emptyset$ for almost all~$\uF$. 
The inequality $\dim R_\lambda(\uF)\geq \dim W -|\lambda|$ follows from \cite[Thm.~17.24]{harr:92}.
\endProofof

\subsection{Expressing transversality}\label{se:trans}

In this section we conclude the proof of Proposition~\ref{pro:projchar}.
We consider input data of the form $(f,n,m,\mu,\uF,x)$
where $f=(f_1,\dots,f_r)$ is a sequence of homogeneous polynomials in $\C[X_0,\dots,X_n]$
satisfying the input condition (\ref{eq:input-cond}) for $m\in\N$ and $x$ is
in the projective zero set $V'$ of these polynomials.
Moreover, $\uF$ is a flag in $\Flag$ encoded by a matrix $a\in\C^{n\times(n+1)}$ and
$\mu=(\mu_1,\dots,\mu_{m+1})$ is an admissible partition with respect to $n$ and $m$.
Recall from Lemma \ref{le:input} the decomposition $V'=V\cup W$, where $V$ is smooth of pure dimension $m$ and
$\dim W<m$.

Let $u\in \C^{\infty}$ be an encoding of $(f,n,m,\mu)$, let $a\in \C^{\infty}$ be an encoding of $\uF$
and define the relation $\trans \subseteq \C^{\infty}\times \C^{\infty}\times \C^{\infty}$ by
\begin{equation*}
  \trans(u,a,x) :\Longleftrightarrow \big(x\in V \wedge \varphi(x)\in e_\mu(\uF)
  \Longrightarrow \varphi\pitchfork_x e_{\mu}(\uF)\big),
\end{equation*}
where $\varphi$ is the Gauss map of $V$.

\begin{lemma}\label{le:trans-ph}
The relation $\trans$ is decidable in polynomial time by a constant-free
machine over~$\C$.
\end{lemma}

Before going into the proof, we recall some facts concerning the 
manifold structure and cell decomposition of Grassmannians. 
For a comprehensive account, we refer to \cite[III.9]{fult:97} and \cite{mani:01}.

Dual to our usual encoding $a\in\C^{n\times(n+1)}$ of a flag $\uF\in \Flag$ 
(where the $F_i$ are zero sets of row forms of~$a$), 
we can represent the flag $\uF$ by a basis $\ell=(\ell_0,\dots,\ell_n)$ of $\C^{n+1}$ 
such that $F_{i}$ is spanned by $(\ell_0,\dots,\ell_{i})$ for $0\leq i\leq n$. 
Clearly, this basis is uniquely determined by $\uF$ up to scaling 
and can be computed from $a$ in polynomial time.

Let $\mu$ be an admissible partition and let $\sigma$ denote the associated sequence 
$0\le\sigma_0<\dots <\sigma_{m}\le n$ defined by $\sigma_i:=n-m+i-\mu_{i+1}$. 
To a fixed basis $\ell$ and $\mu$ we assign the Schubert cell 
$e_\mu:= e_\mu(\ell):=e_\mu(\uF)$ according to~(\ref{eq:def-scell}). 
(To ease notation, we will usually drop the dependence on $\ell$.) 
It is not hard to see that every subspace $A$ in $e_\mu$ has a unique basis, 
that can be represented with respect to the basis~$\ell$
by the rows of an $(m+1)\times (n+1)$ row echelon matrix, which 
has a~$1$ at the intersection of the $i$-th row with the $\sigma_i$-th column, and 
zeros in the $i$-th row to the right of this position as well as zeros in the $\sigma_i$-th column below this position, 
for all $0\leq i\leq m$.
In the case $m=3,n=7$, $\mu=(3,1,0)$, $\sigma=(1,4,6,7)$ such an echelon matrix looks as follows:
\begin{equation}\label{eq:echelon}
\begin{pmatrix}
* & 1 & 0 & 0 & 0 & 0 & 0 & 0 \\
* & 0 & * & * & 1 & 0 & 0 & 0 \\
* & 0 & * & * & 0 & * & 1 & 0 \\
* & 0 & * & * & 0 & * & 0 & 1 
\end{pmatrix}.
\end{equation}

In order to describe a covering of $\Gr(m,n)$ in terms of affine charts, consider for fixed $\ell$  
the subspaces $L_\mu$ and $\overline{L}_\mu$ of $\C^{n+1}$ spanned by 
$\ell_{\sigma_0},\dots,\ell_{\sigma_{m}}$ and $\{\ell_j \mid j\not\in \{\sigma_0,\ldots,\sigma_m\}\}$, 
respectively.    
We define $U_\mu:=U_\mu(\ell)\subseteq\Gr(m,n)$ as the set of $(m+1)$-dimensional
subspaces $A\subseteq \C^{n+1}$ whose projection to the subspace $L_\mu$ along $\overline{L}_\mu$ 
is an isomorphism.
The open sets $U_\mu$ form an open cover of $\Gr(m,n)$. 
By identifying $A\in U_\mu$ with the graph of a linear map from $L_\mu$ to $\overline{L}_\mu$, 
we get an isomorphism 
\begin{equation}\label{eq:chartmap}
\alpha_\mu\colon U_\mu \xrightarrow{\sim} 
 \Hom(L_\mu,\overline{L}_\mu)\xrightarrow{\sim} \C^{(n-m)\times (m+1)},
\end{equation}
where the last isomorphism maps an element of $\Hom(L_\mu,\overline{L}_\mu)$ to its matrix representation 
with respect to the bases defined by $\ell$.
The matrix $\alpha_\mu(A)$ is obtained from the echelon matrix in~(\ref{eq:echelon}) 
by removing all the $\sigma_i$-columns (thus removing a unit matrix of size $m+1$) and transposing. 
Taking this into account, we see that $e_\mu \subset U_\mu$ and 
that the image of $e_\mu$ under $\alpha_\mu$ can be described as follows:
\begin{equation}\label{eq:image-cell}
  \alpha_\mu(e_\mu)=\{ (a_{ij}) \in \C^{(n-m)\times (m+1)}\mid 
  \mbox{ $a_{ij}=0$ for $j\geq \sigma_i-i, 0\leq i\leq m,0\le j< n-m$}\}.
\end{equation}
In particular, $\alpha_\mu(e_\mu)$ is a linear subspace of $\C^{(n-m)\times(m+1)}$. 

\Proofof{Lemma \ref{le:trans-ph}}
Assume that $x\in V$ and $\varphi(x)\in e_\mu(\ell)$. 
The claim is that the transversality condition 
\begin{equation}\label{eq:transx}
 T_{\varphi(x)}\Gr(m,n)=d_x\varphi(T_xV)+T_{\varphi(x)}e_\mu(\ell).
\end{equation}
can be checked in constant-free polynomial time over $\C$.

In order to simplify notation, we will identify~$V$ with its affine cone~$\hat{V}$, 
$x$ with an affine representative $\hat{x}$, and the Gauss map $\varphi$ with the 
corresponding morphism $\hat{\varphi}\colon \hat{V}-\{0\}\rightarrow \Gr(m,n)$.
This causes no problem, since $d_x\varphi(T_xV)=d_{\hat{x}}\hat{\varphi}(T_{\hat{x}}\hat{V})$. 

Given a basis~$\ell$ and a partition~$\mu$, we represent $e_\mu= e_\mu(\ell)$ and
the tangent spaces $T_A\Gr(m,n)$ and $T_Ae_\mu$  for $A\in e_\mu$
by means of the chart $ \alpha_\mu$ defined in~(\ref{eq:chartmap}).
Around $x$ we extend the Gauss map $\varphi$ into the chart considering 
\begin{equation*}
 \varphi_\mu\colon V\cap \varphi^{-1}(U_\mu)\xrightarrow{\varphi} 
  U_\mu\xrightarrow{\alpha_\mu}\C^{(n-m)\times (m+1)}
\end{equation*}
In this light, Equation~(\ref{eq:transx}) translates into
\begin{equation*}
 \C^{(n-m)\times (m+1)}=d_x\varphi_\mu(T_xV)+\alpha_\mu(e_\mu).
\end{equation*}
Equation~(\ref{eq:image-cell}) gives an explicit and simple description of $\alpha_\mu(e_\mu)$.
It remains to find a suitable description of $d_x\varphi_\mu (T_xV)$. 

After a linear coordinate transformation, we may assume that 
$L_\mu =\C^{m+1}\times 0$ and $\overline{L}_\mu = 0\times\C^{n-m}$.  
Thus without loss of generality, we assume that 
$X_0,\ldots,X_n$ are coordinates adapted to the decomposition 
$\C^{n+1}=L_\mu\oplus\overline{L}_\mu$. 

Locally around the point~$x$, the variety $V\subseteq\C^{n+1}$ is given as the zero set 
of the polynomials $f_1,\ldots,f_r$. 
Our assumption $\varphi(x)\in e_\mu$ means that $T_xV$ lies in $e_\mu$ 
and thus in $U_\mu$.
This implies that the matrix 
$(\frac{\partial f_s}{\partial X_{t}}(x)))_{1\leq s\leq r,m< t\leq n}$
has rank $n-m$. After a permutation, we may assume that 
$(\frac{\partial f_s}{\partial X_{t}}(x)))_{1\leq s\leq n-m,m< t\leq n}$
is invertible.
It will be convenient to use the abbreviations 
$X':=(X_0,\dots,X_m)$ and $X'':=(X_{m+1},\dots,X_n)$. 

By the implicit function theorem there are analytic functions 
$h_1,\dots,h_{n-m}$ in $X'$ such that in a neighborhood of~$x$, the variety~$V$ 
is the graph of the analytic function $h:=(h_1,\dots,h_{n-m})$ defined on a neighborhood of~$x'$. 
In particular, $x=(x',h(x'))$. 
From this we obtain the following description of the Gauss map: 
$$
 \varphi_\mu(X',h(X')) = 
 \Big(\frac{\partial h_s}{\partial X_i}(X')\Big)_{1\leq s\leq n-m,0\leq i\leq m}
 \in\C^{(n-m)\times (m+1)}.
$$
Hence the vector space $d_x\varphi_\mu(T_xV)$ is spanned by the matrices
$$
 \Big(\frac{\partial^2 h_s}{\partial X_i\partial X_j}(x')\Big)_{1\leq s\leq n-m,0\leq i\leq m}
$$
for $0\le j\le m$. 
It remains to show that these matrices can be computed in constant-free polynomial time over $\C$. 
We remark that in the case of a hypersurface ($m=n-1$), this matrix just describes the 
second fundamental form of $V$ at~$x$. 

By taking the derivative with respect to $X_i$ of $f_s(X',h(X'))= 0$, we obtain 
\begin{equation}\label{eq:1st-deriv}
 \frac{\partial f_s}{\partial X_i}(X',h(X')) + 
 \sum_{t=m+1}^n \frac{\partial f_s}{\partial X_t}(X',h(X'))\frac{\partial h_t}{\partial X_i}(X') = 0
\end{equation}
for $1\le s\le n-m, 0\le i \le m$. 
From this, $\frac{\partial h_t}{\partial X_i}(x')$ can be computed by inverting the 
matrix $(\frac{\partial f_s}{\partial X_t}(x))$. 
By taking the derivative of Equation~(\ref{eq:1st-deriv}) with respect to~$X_j$ 
for $0\le j\le m$ we get
\begin{equation*}
 \frac{\partial^2 f_s}{\partial X_i\partial X_j} 
 +2\sum_{t> m}\frac{\partial^2 f_s}{\partial X_t\partial X_j} \frac{\partial h_t}{\partial X_i} 
 +\sum_{t,k > m}\frac{\partial^2 f_s}{\partial X_t\partial X_k}\frac{\partial h_t}{\partial X_i}\frac{\partial h_k}{\partial X_j}  
 +\sum_{t>m}\frac{\partial f_s}{\partial X_t}\frac{\partial^2 h_t}{\partial X_i\partial X_j} = 0 .
\end{equation*}
From this, the desired second order derivatives $\frac{\partial^2 h_t}{\partial X_i\partial X_j}(x')$  
can be computed by inverting the matrix $(\frac{\partial f_s}{\partial X_t}(x))$. 
This finishes the proof.
\endproofof

{\small

}
\end{document}